\documentclass[a4paper,fleqn,usenatbib]{mnras}

\usepackage[T1]{fontenc}
\usepackage{ae,aecompl}
\usepackage{pdflscape}

\usepackage{graphicx}	
\usepackage{amsmath}	
\usepackage{amssymb}	

\usepackage{txfonts}

\title[Products of stellar evolution in M67]{Observing the products of stellar evolution in the old open cluster M67 with APOGEE}

\author[Bertelli Motta et al.]{
Clio Bertelli Motta,$^{1}$\thanks{E-mail: cbertelli@ari.uni-heidelberg.de}
Maurizio Salaris,$^{2}$
Anna Pasquali,$^{1}$
Eva K. Grebel$^{1}$
\\
$^{1}$Astronomisches Rechen-Institut, Zentrum f\"ur Astronomie der Universit\"at Heidelberg, M\"onchhofstr. 12-14, 69120 Heidelberg, Germany\\
$^{2}$Astrophysics Research Institute, Liverpool John Moores University, 146 Brownlow Hill, Liverpool L3 5RF, UK\\
}

\date{Accepted 2016 December 12. Received 2016 December 9; in original form 2016 August 9}

\pubyear{2016}

\begin{document}
\label{firstpage}
\pagerange{\pageref{firstpage}--\pageref{lastpage}}
\maketitle

\begin{abstract}
Recent works have shown how the [C/N] ratio in stars after the first dredge-up (FDU) can be used as an age estimator in virtue of its dependence on stellar mass. For this purpose, precise predictions of the surface chemical composition before and after the mixing takes place in the convective envelope of subgiant stars are necessary. Stellar evolution models can provide us with such predictions, although a comparision with objects of known age is needed for calibration. Open clusters are excellent test cases, as they represent a single stellar population for which the age can be derived through, e.g., isochrone fitting. In this study, we present a detailed analysis of stars belonging to the well-known open cluster M67 observed by the APOGEE survey in the twelfth data release of the Sloan Digital Sky Survey and whose chemical properties were derived with the ASPCAP pipeline. We find that the [C/N] abundance of subgiant branch stars is overestimated by $\sim0.2$ dex due to an offset in the determination of the [N/Fe] abundance. Stars on the red giant branch and red clump are shown not to be affected by this offset. We derive $\text{[C/N]}_{FDU}=-0.46\pm 0.03$ dex, which poses a strong constraint on calibrations of $\text{[C/N]}_{FDU}$
as age indicator. We also do not find any clear signature of additional chemical mixing processes that set in after the red giant branch bump. The results obtained for M67 indicate the importance of conducting high-resolution spectroscopic studies of open clusters of different ages in order to establish an accurate age-dating method for field stars.
\end{abstract}




\section{Introduction}
 Age-dating of field stars is crucial in order to reconstruct the evolutionary history of the Galaxy. Determining the age distribution throughout the Milky Way requires observations and age-dating of stars brighter than the main-sequence turn-off, in evolutionary phases that cover a large range of the Galactic evolutionary history. 
Red giant branch (RGB) stars are particularly useful for this purpose, for they are bright and span an age range between $\sim$1~Gyr 
and the age of the Universe, and indeed modern spectroscopic surveys \citep[e.g., the {\sl Gaia}-ESO and the Apache Point Observatory Galactic Evolution Experiment --APOGEE-- surveys, see ][]{gilmore2012, holtzman2015} provide surface gravity ($\log g$), effective temperature ($T_{eff}$), and photospheric abundances of several chemical elements for large samples of Milky Way RGB stars.
 
Age estimates of RGB stars are however particularly challenging, since small uncertainties in $T_{eff}$ (at fixed $\log g$) cause large uncertainties in the stellar mass, hence their age, if asteroseismic constraints on the mass of the object are not available. 
To improve RGB star age estimates, \citet{masseron2015} employed as age diagnostic the abundance ratio ${\rm [C/N]_{FDU}}$ measured after the completion of the FDU and at magnitudes fainter than the RGB bump luminosity, where observations reveal the onset of an extra mixing process that continues until the tip of the RGB \citep[and beyond -- see, e.g.,]
[for more details]{salaris2015}. 

Stellar model calculations show that at the end of the main sequence phase the outer convection zone progressively engulfs deeper regions, reaching layers that had been partially processed by H-burning during the main sequence phase.
In these layers --whether or not the CNO-cycle was the main energy-generation mechanism-- the C and N abundances had enough time to attain the CN-cycle equilibrium values, implying an increase of N and a decrease of C with respect to an initial scaled-solar or
$\alpha$-enhanced metal mixture.

As a consequence, convection dredges N-enriched and C-depleted matter to the surface (FDU),
lowering the surface [C/N] ratio compared to the initial value. This decrease of [C/N] after the FDU is mass dependent, because when the stellar mass increases, surface convection at the FDU engulfs a larger fraction of the total stellar mass,  causing a larger decrease of the surface [C/N] with increasing RGB mass -- hence with decreasing age of the stellar population.

\citet{martig2016} provided a semi-empirical calibration of ${\rm [C/N]_{FDU}}$-chemical composition-age relations, based on a sample of field stars with also asteroseismic mass estimates, for which the age was determined using mass-age relations from theoretical stellar models. The accuracy of the ages determined by means of this relation is about 40\% \citep{martig2016}.
 
On the theoretical side, \citet{salaris2015} presented and discussed a theoretical calibration of ${\rm [C/N]_{FDU}}$ as a function of metallicity and age, which allows one to age-date RGB stars with an internal accuracy of about $\sim$15\%.
However, as shown by these authors, different sets of theoretical models give very different results in terms of the 
${\rm [C/N]_{FDU}}$-age relation at a given metallicity, especially for ages below 10~Gyr. 
For this reason, tests of the theoretical ${\rm [C/N]_{FDU}}$-age relationships in open clusters with well-determined metallicity 
and age are absolutely crucial. 

In this paper we consider the old open cluster M67 (or NGC 2682). Located at 800-900 pc from the Solar system (see, e.g., \citealt{sarajedini2009}), M67  is one of the most interesting and best studied examples among Galactic open clusters.  With an age of 3.5-4 Gyr (see, e.g., \citealt{sarajedini2009,bellini2010,kharchenko2013}), M67 is very old for an open cluster. Moreover, its chemical composition was shown to be more similar to that of the Sun than most other nearby field stars. In particular, the solar twin M67-1194 is considered to be the star most similar to the Sun observed so far. These findings led to the hypothesis that the Sun might have formed inside M67, although in this scenario, the protoplanetary disc or planetary system of the Sun should have been disrupted during the ejection process (see \citealt{pichardo2012}). 

For M67, spectroscopic values of ${\rm [C/N]_{FDU}}$ from \citet{gilroy1991} were compared by \citet{salaris2015} with their theoretical ${\rm [C/N]_{FDU}}$-age relation for the cluster metallicity. However, the uncertainties in the empirical data were very large and the comparison did not set any strong constraint on such relation.
Here we use chemical abundances of C, N, O, and Fe taken from APOGEE published as part of SDSS-III data release 12 (DR12, 
\citealt{alam2015}) for a large sample of M67 stars covering evolutionary phases ranging from the subgiant branch to the red clump. These data do not only provide a better estimate of ${\rm [C/N]_{FDU}}$ --hence a stronger constraint on theory -- but also allow one to follow the [C/N] variation during the FDU, and potential additional mixing after the RGB bump phase. 

The plan of the paper is as follows. Section~2 describes the spectroscopic data and the cluster membership analysis, 
whilst Section~3 compares the spectroscopic abundances with results from theoretical models, 
and the ${\rm [C/N]_{FDU}}$-age relation by \citet{salaris2015}. A discussion and conclusions close the paper.

\section{Data}
\label{sec:data}

In this work, we analyse stellar spectra observed within the Apache Point Observatory Galactic Evolution Experiment \citep[APOGEE, see][]{majewski2015} as part of the Sloan Digital Sky Survey III \citep[SDSS-III, DR12, see][]{eisenstein2011,alam2015}. The APOGEE spectrograph is mounted on the 2.5m Sloan telescope and operates in the near-infrared range from 1.51 to $1.7 \mu$m with a spectroscopic resolution of $R \sim 22,500$.  Stellar parameters such as $T_{eff}, \log g, \text{[M/H], [C/M], [N/M], and [}\alpha\text{/M]}$, as well as the abundances of several elements are derived with the APOGEE Stellar Parameters and Chemical Abundances Pipeline (ASPCAP) \citep[see][]{garciaperez2015,holtzman2015}.

\subsection{Membership analysis}
\label{sec:meth}

For an investigation of the chemical composition of stars in open clusters and especially for the study of possible inhomogeneities arising from stellar evolution, a reliable determination of the membership probability of each star is essential. In fact, only if we can exclude contamination from field stars with (almost) absolute certainty, we can reasonably assume that potential variations in the chemical abundances of cluster stars are a product of either stellar evolution or of the star formation history within the cluster.

\begin{figure*}
	\includegraphics[scale=0.3]{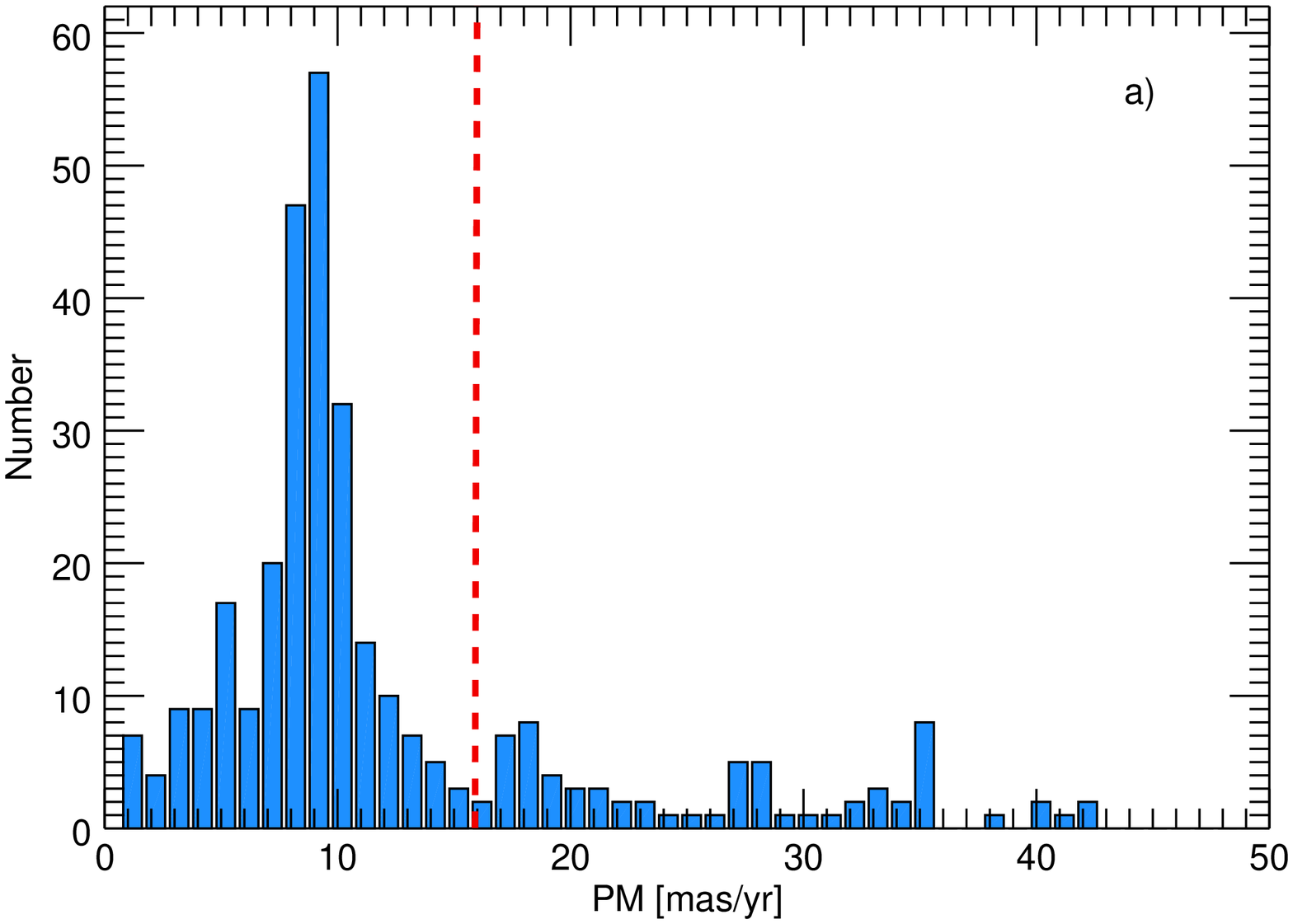}
	\includegraphics[scale=0.3]{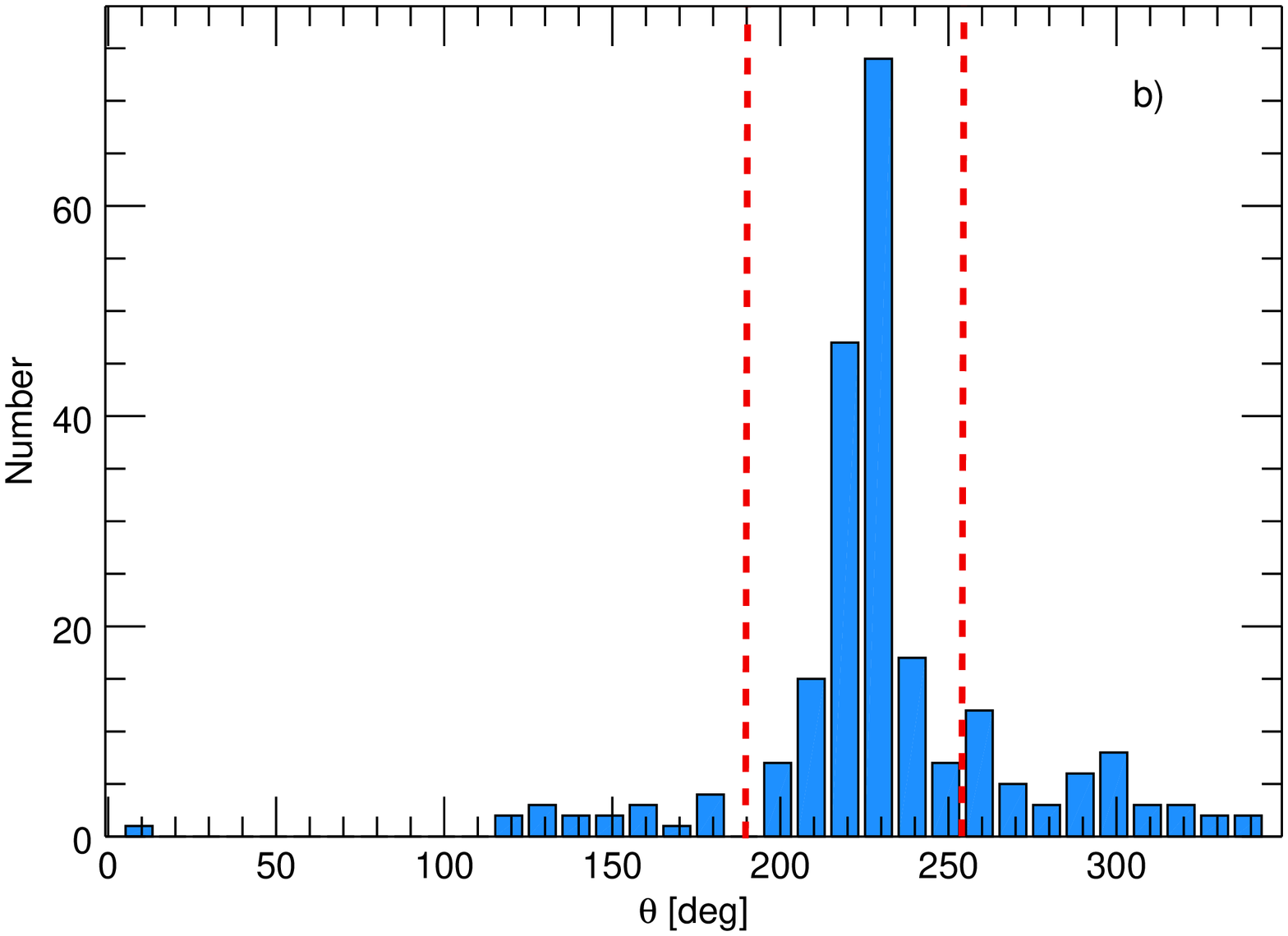}
	\includegraphics[scale=0.3]{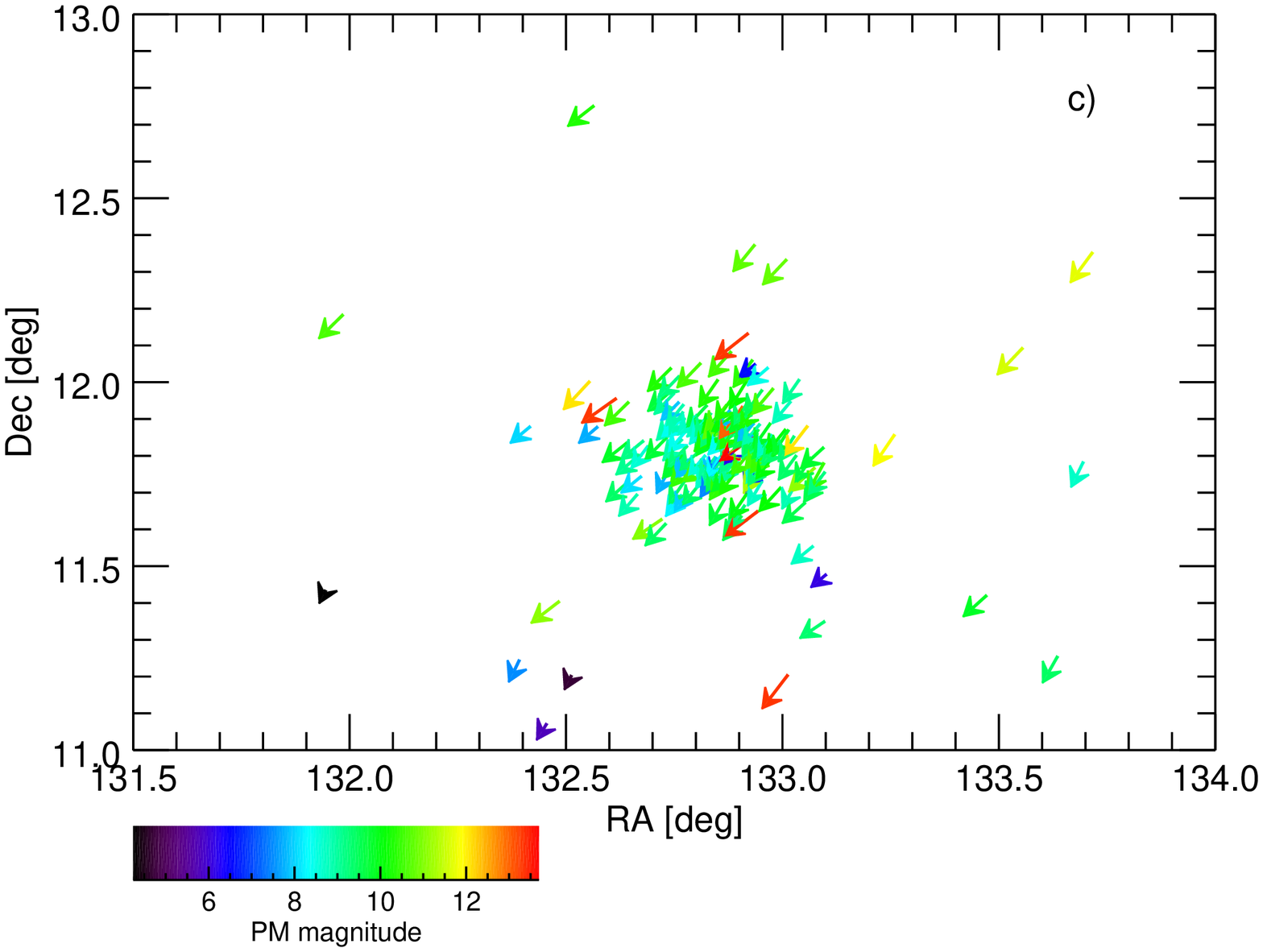}
	\includegraphics[scale=0.3]{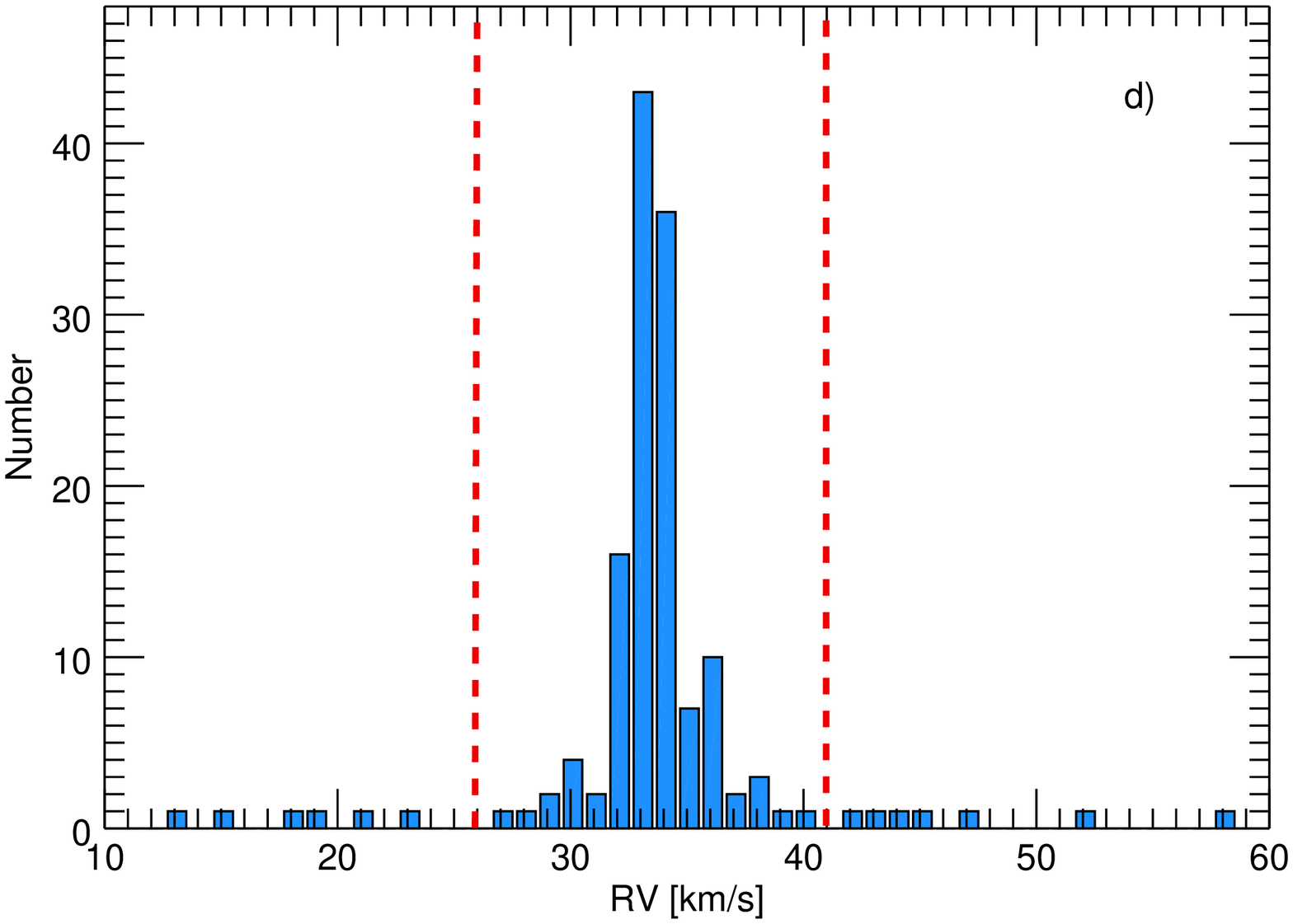}
	\includegraphics[scale=0.3]{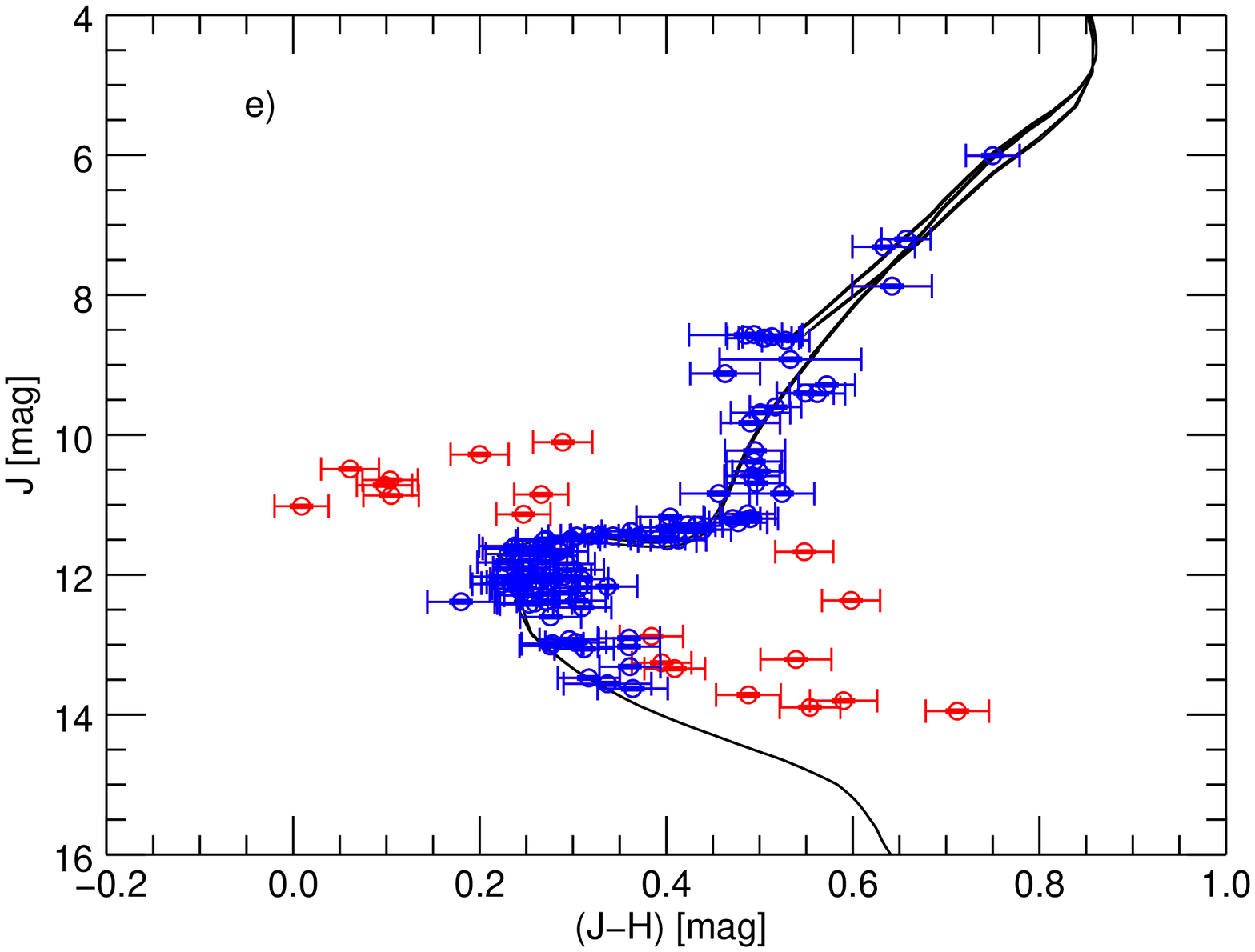}
	\includegraphics[scale=0.3]{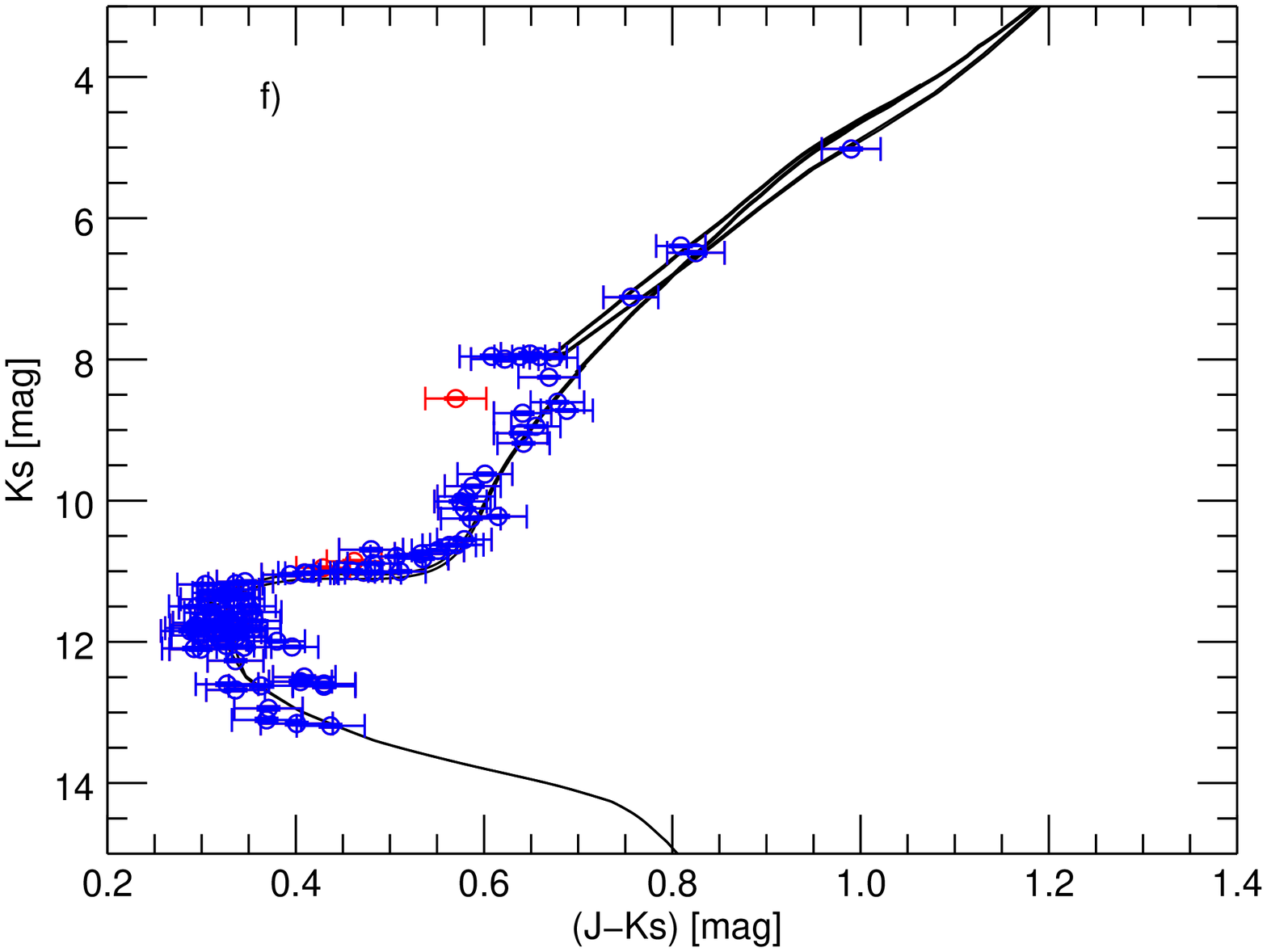}
	\includegraphics[scale=0.3]{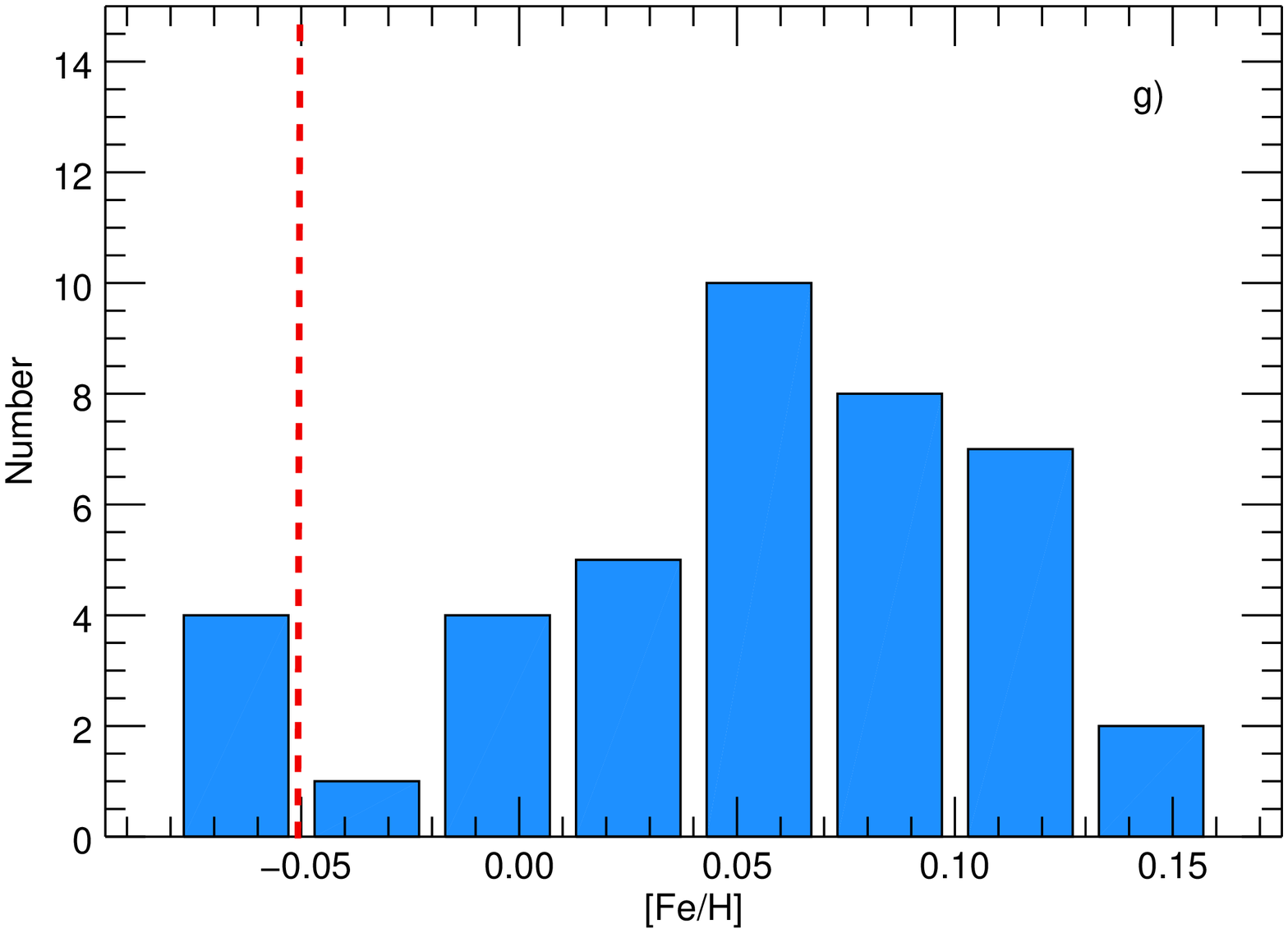}
	\caption{The membership analysis step by step. \textit{Panel a}: histogram of $PM_{tot}$ (mean error $PM_{err}\sim1.8$ mas/yr); \textit{panel b}: histogram of the proper motion angle ($\theta_{err}\sim9.88$ deg); \textit{panel c}: representation of the proper motion vector of the selected stars in the RA-Dec plane; \textit{panel d}: histogram of the radial velocities ($RV_{err}\sim0.05$ km/s); \textit{panel e} and \textit{f}: colour-magnitude diagram for $J$ vs. $(J-H)$ and $K_s$ vs. $(J-K_s)$, the rejected stars are highlighted in red; \textit{panel g}: histogram of the [Fe/H] abundances of the selected stars ($\text{[Fe/H]}_{err}\sim0.03$ dex). The red dashed lines represent the interval taken into account for the computation of the mean value and standard deviation of each distribution.}
	\label{fig:analysis}
\end{figure*}

In order to select likely member stars of M67, we developed a code that analyses different properties of the candidate members sample.  In the following, we illustrate in detail the membership analysis pipeline.
\begin{itemize}
	\item [\textbf{1.}] The first step consists in a crossmatch of the APOGEE stars in the SDSS DR12 with the central coordinates of M67, which we retrieve 	from \citet{kharchenko2013}. All stars are selected that lie within the cluster's outermost radius (1.03 deg), defined by  \citet{kharchenko2013} as the distance from the centre of the cluster at which the density of stars cannot be distinguished from that of the field. 
	\item [\textbf{2.}] We expect that the member stars of a given open cluster share the same tangential velocities (as inferred from their proper motions in RA and Dec) and radial velocities (RV). Thus, after the spatial constraints set by step 1., we address the tangential velocities of the stars. In particular, we are interested in the absolute value of the proper motion (PM) vector, which we derive through the vectorial sum of the PM components in RA and Dec provided by the PPMXL catalogue  \citep{roeser2010}, including the corrections by \citet{vickers2016}. The following relation holds for the absolute value of the PM:
	\begin{equation}
		PM_{tot}=\sqrt{PM_{RA}^2+PM_{Dec}^2}.
		\label{eq:pm}
	\end{equation} 
	Plotting a histogram of the resulting $PM_{tot}$ for all stars within the radius of the cluster, we expect the members of the cluster to show a distribution peaked at the mean $PM_{tot}$ of the cluster and whose width is given by the PM errors from the PPMXL catalogue and by the internal scatter due to the progressive dissolution of the cluster. In order to exclude obvious outliers from the calculations, we select a first range by eye (red lines in the plots in Fig.~\ref{fig:analysis}, panel a). We then calculate the weighted mean $PM_{tot}$ and its standard deviation within this range. All stars lying within $2 \sigma$ from the mean are selected and considered in the next step of the analysis. 
	
	Due to the very large $PM_{tot}$ errors for some of the stars in the PPMXL catalogue that we do not want to include in our sample, we calculate the mean error for the above selection and exclude stars whose $PM_{tot}$ error is larger than the mean error ($\sim1.8$ mas/yr) $+1\sigma$ ($\sim1.2$ mas/yr).
	\item [\textbf{3.}] In step 2, we consider only the absolute value of the $PM_{tot}$ and thus lose information about the direction of the $PM_{tot}$ vector. 	Therefore, it might happen that stars having by chance the same absolute $PM_{tot}$ as the average proper motion of the cluster, but moving in a different direction, are considered members of the cluster. In order to correct for this effect, we consider in the next step the distribution of the angles of the $PM_{tot}$ vector (see Figure~\ref{fig:analysis}, panel b). Also here, we calculate the mean weighted by the errors (mean error $\sim9.88$ deg) and the standard deviation and select all objects within $2\sigma$ from the mean. 
	The result of this selection can be seen in Figure~\ref{fig:analysis}, panel c: all stars move in the same direction in the RA-Dec plane, with proper motion vectors of the same length (within the errors).
	\item [\textbf{4.}] In the next step we repeat the same procedure as in step 2 and 3, but with the RV of the stars resulting from  the ASPCAP pipeline (see Figure~\ref{fig:analysis}, panel d). Given that the typical errors of the ASPCAP-derived RV for stars in the radius of M67 are very small ($\sim0.05$km/s), we do not put any constraint on these errors, contrarily to step 1.
	\item [\textbf{5.}] Once we have made sure that the kinematic properties of our selection of stars are consistent with them belonging to the same kinematic group, we proceed with the photometric criteria. For all selected stars we can retrieve 2MASS magnitudes. We use for the distance modulus $(m-M)_0=9.64$ mag, for the reddening E(B-V)=0.023 mag and for the age estimate the range between 3.75 and 4 Gyr \citep[from][]{bellini2010} to calculate the corresponding BaSTI isochrones (see \citealt{pietrinferni2004}) with a metallicity of $\text{[Fe/H]=0.06}$ dex, including core overshooting during the main sequence. We then compare the isochrones reddened according to the \citet{cardelli1989} extinction law with the 2MASS magnitudes on a colour-magnitude diagram. 
	We exclude the objects that do not cross any of the two isochrones corresponding to an age of 3.75 and 4 Gyr within three times their error in magnitude and colour. We repeat this step in the colour-magnitude diagrams described by $J$ vs $(J-H)$ and $K_s$ vs $(J-K_s)$ (see Figure~\ref{fig:analysis}, panel e and f). 
	
	We should note that by doing this we exclude blue stragglers that are probably members of M67. For the purpose of this study this is not relevant, since blue stragglers due to their anomalous origin cannot be taken into account when investigating the effects of stellar evolution on the surface abundances of [C/N]. 
	\item [\textbf{6.}] As a final step, we consider that stars belonging to a given open cluster, even in case of inhomogeneities in the abundances of certain elements, are expected to share the same [Fe/H] abundance. Thus, we repeat the same procedure as in steps 2.-4. for the iron abundance [Fe/H] calculated with the pipeline ASPCAP, and consider all stars within $2\sigma\approx0.1$ dex from the mean [Fe/H] to be members of the cluster (see Figure~\ref{fig:analysis}, panel g). Typical [Fe/H] errors for this selection are $\sim0.03$ dex.
	
	We point out that the DR12/ASPCAP results for dwarf stars have not been calibrated due to the difficulties of the pipeline in analysing stars with high $\log(g)$. Therefore, in step 6 we exclude potential cluster members on the main sequence and turn-off of M67 from our selection because they do not have entries in the DR12/ASPCAP catalogue of calibrated abundances. For the present study this is not very relevant, since our purpose is not to obtain a complete sample of M67 members but rather to select a number of highly probable members that allow us to investigate the effects of stellar evolution on the chemical composition of stars. Therefore, the sample resulting from our pipeline contains stars from the subgiant and giant branch, which is perfect in order to analyse the effects of the FDU on the surface abundances of [C/N]. 
	
\end{itemize}

\begin{figure}
	\includegraphics[scale=0.39]{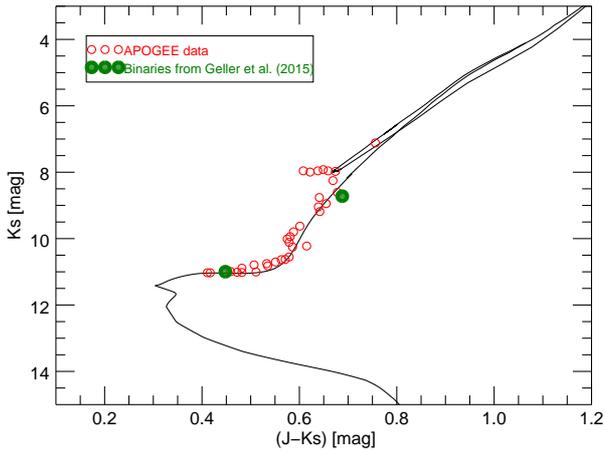}
	\caption{Our selection of members for M67 plotted on the BaSTI isochrone of age 3.75 Gyr. The green circles represent the members labeled as binaries 
in \citet{geller2015}. These stars are excluded from all further investigations.}
	\label{fig:cmd_bin}
\end{figure}

After the cluster members selection, we cleaned the sample from binaries known from the literature. We matched our sample with those of \citet{yakut2009}  and \citet{geller2015}. We did not find any stars in common with \citet{yakut2009}, but we did find two binaries from \citet{geller2015} among our members, as shown in Fig.~\ref{fig:cmd_bin}. We exclude these two stars from our sample, since being unresolved binaries, we cannot use their abundances to investigate the effects of stellar evolution on the surface chemistry of single cluster stars.

Averaging over the resulting sample of 34 stars distributed over the subgiant branch, the lower red giant branch and the red clump, we find M67 to have an average radial velocity of $RV=33.806\pm0.528$ km/s, a proper motion in $\text{RA}\times\cos(\text{Dec})$ of $PM_x=-5.906\pm0.951$ mas/yr and in Dec of $PM_y=-7.175\pm1.268$ mas/yr (using the corrected proper motions from \citealt{vickers2016})\footnote{Repeating the membership analysis with the non-corrected proper motions from PPMXL we find $PM_x=-7.754\pm0.952$ mas/yr and $PM_y=-5.613\pm1.276$ mas/yr.}, consistent with the literature (see \citealt{yadav2008,geller2015} and \citealt{bellini2010b}). For the metallicity, we obtain $\text{[Fe/H]}=0.08\pm0.04$ dex, slightly higher than other literature values \citep[see, e.g.,][]{tautv2000,shetrone2000,yong2005,randich2006,pace2008,onehag2014}, but still consistent with them within the errors. The cluster-wide parameters of M67 have been summarized in Table~\ref{tab:cluster_par}.

\begin{figure*}
	\includegraphics[scale=0.39]{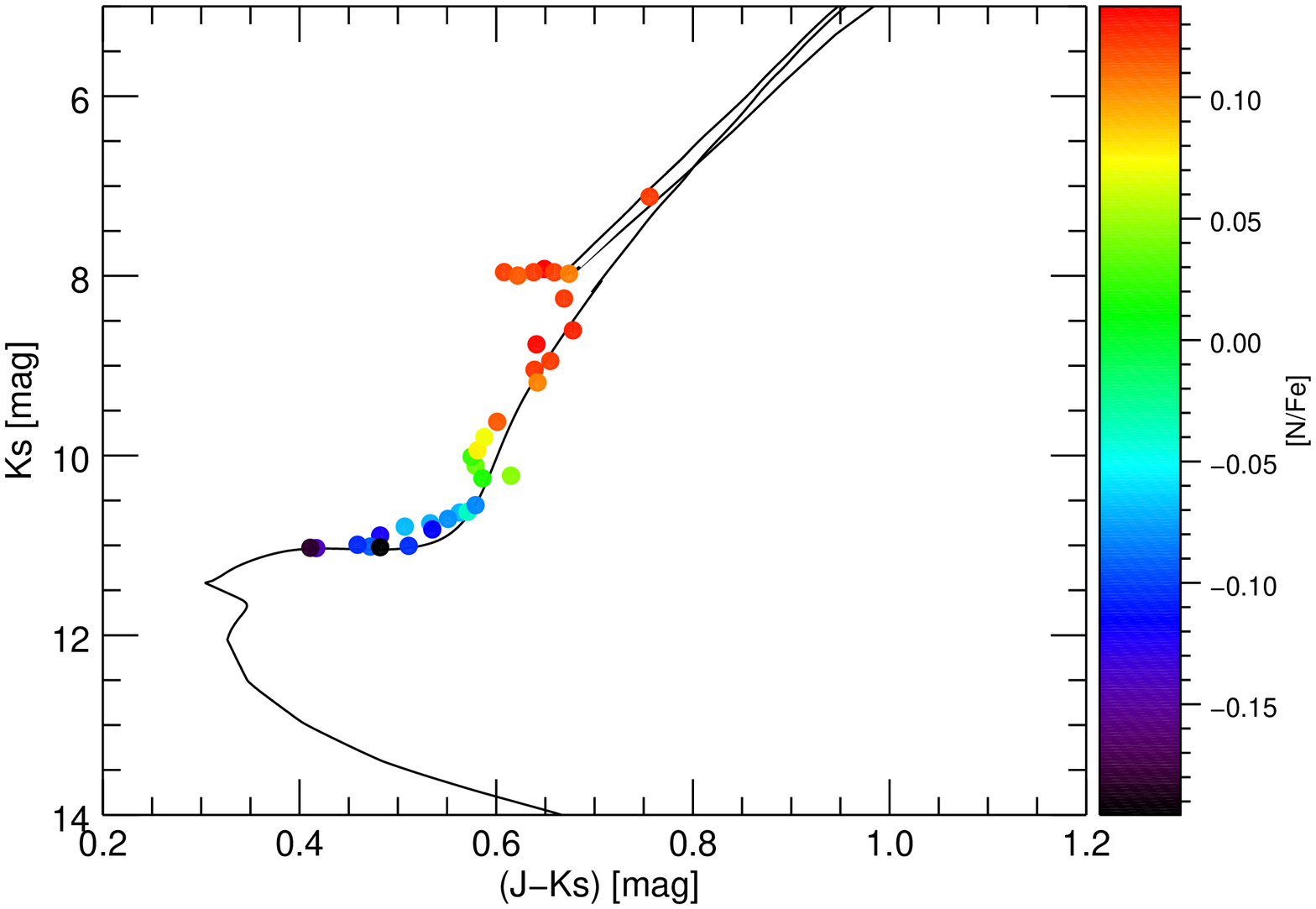}
	\includegraphics[scale=0.39]{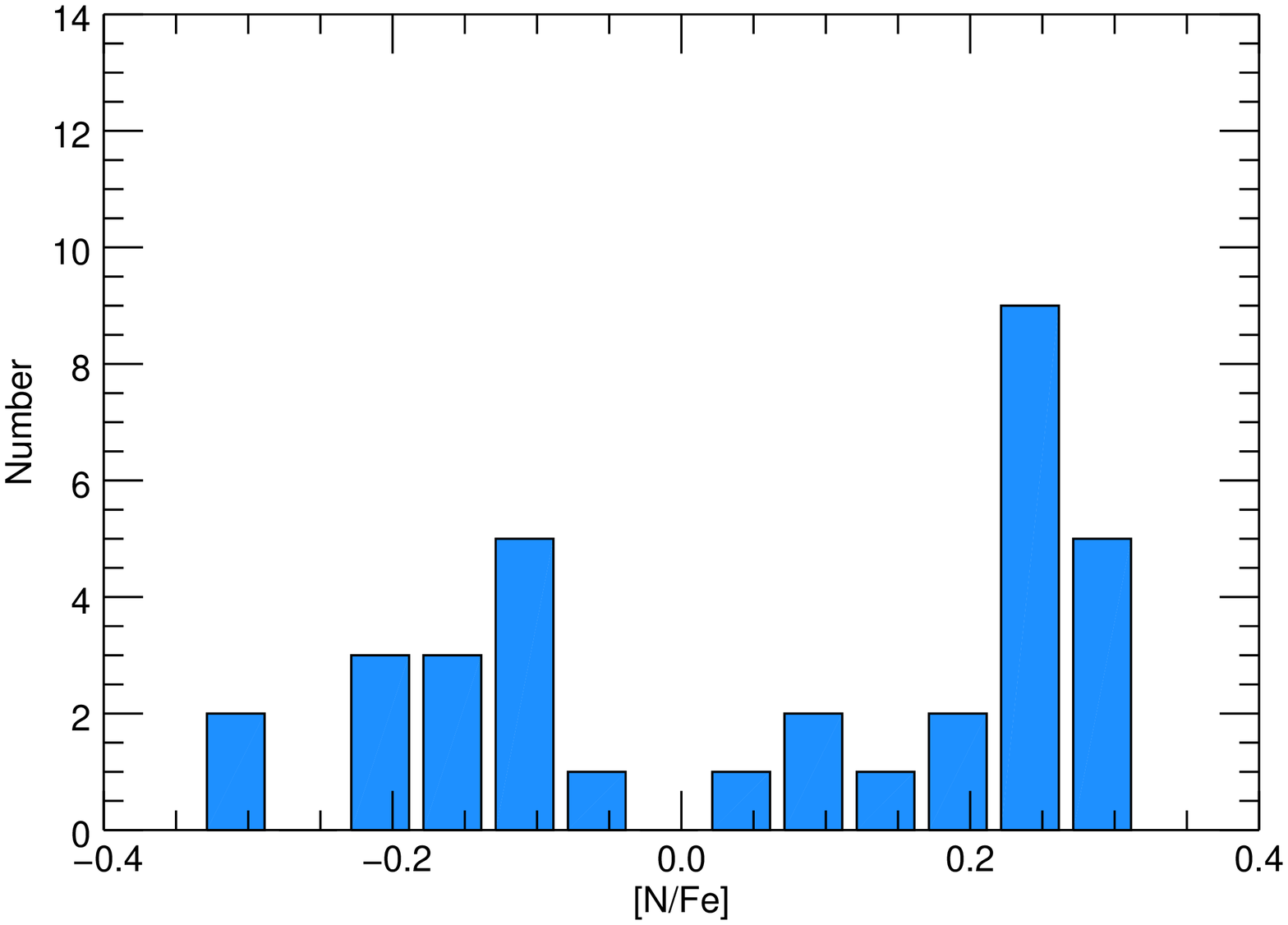}
	\caption{\textit{Left panel:}Colour-magnitude diagram of the member stars of M67with an overplotted 3.75 Gyr isochrone. The symbols representing the stars are colour-coded with their [N/Fe] abundance. The effect of the first dredge-up can be seen clearly at the transition between the subgiant and the giant branch. The N produced by the CNO cycle is brought to the surface, thus enhancing its abundance. \textit{Right panel:} Histogram showing the [N/Fe] abundance distribution of the member stars of M67 (mean error $\text{[N/Fe]}_{err}\sim0.08$ dex). The distribution does not represent the expectations for the chemical composition of the main sequence of an open cluster, i.e. a single peak with a small variation. The abundance distribution of the subgiants and giants shows instead two	peaks in [N/Fe].}	
	\label{fig:ab_iso_nfe}
\end{figure*}

\begin{figure*}
	\includegraphics[scale=0.39]{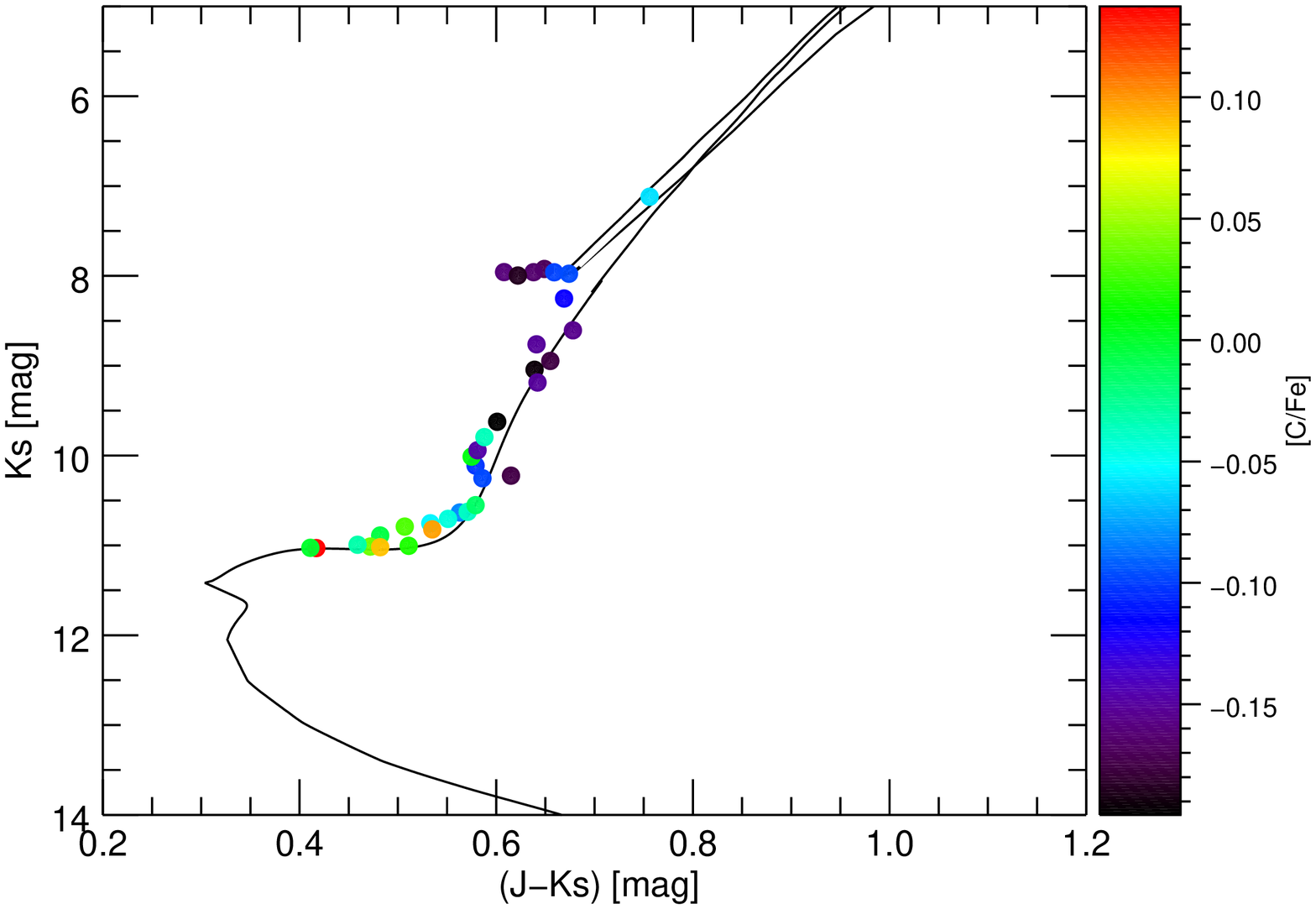}
	\includegraphics[scale=0.39]{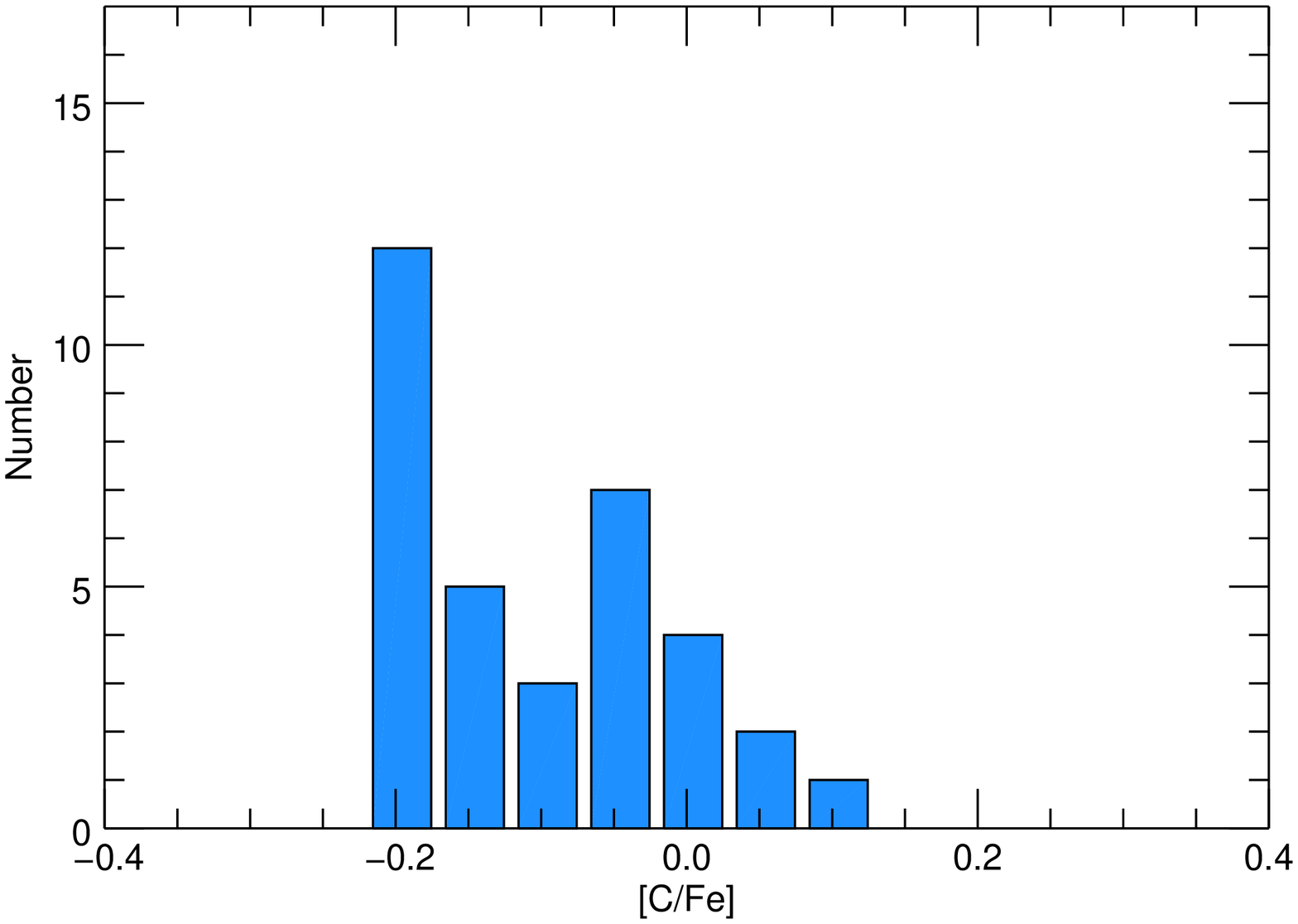}
	\caption{\textit{Left panel:} A 3.75 Gyr isochrone is shown overplotted on the members of M67. The member stars are colour-coded according to their [C/Fe] abundance. In contrast to N, the abundance of C decreases during the CNO cycle, and as a consequence the surface abundance of [C/Fe] after the first dredge-up is depleted. \textit{Right panel:} Histogram of the [C/Fe] abundance of the member stars of M67 (mean error $\text{[C/Fe]}_{err}\sim0.06$ dex). Also in this case, two peaks in the abundance distribution are visible, although not as distinctly as in the case of [N/Fe].}
	\label{fig:ab_iso_cfe}
\end{figure*}

For the stars selected above, we perform an analysis of the elemental abundances affected by stellar evolution, in particular [C/Fe] and [N/Fe] (Tables~\ref{tab:members_par} and ~\ref{tab:members_ab} contain information about the parameters and abundances of the 34 selected member stars). These effects are very well illustrated by Fig.~\ref{fig:ab_iso_nfe} and Fig.~\ref{fig:ab_iso_cfe}. The [C/Fe] and [N/Fe] abundances of the M67 members vary along the isochrone, starting with the main sequence abundances on the subgiant branch and gradually showing the effect of the FDU on the red giant branch. In addition, the [C/Fe] and [N/Fe] distributions exhibit two peaks representing the pre- and post-FDU abundances.

\begin{table*}
	\caption{General parameters of M67. The central coordinates are taken from \citet{kharchenko2013}. The quoted radial velocity, the proper motions, and the metallicity are the mean values for the sample of cluster members selected in this study. The turn-off age range, as well as the reddening and the distance modulus are taken from \citet{bellini2010}.}
	\begin{tabular}{|c|c|c|c|c|c|c|c|c|c|c|c|c|}
		\hline
		\multicolumn{1}{c|}{RA} &
		\multicolumn{1}{c|}{Dec} &
		\multicolumn{1}{c|}{RV} &
		\multicolumn{1}{c|}{e\_RV} &
		\multicolumn{1}{c|}{PM\_x} &
		\multicolumn{1}{c|}{e\_PM\_x} &
		\multicolumn{1}{c|}{PM\_y} &
		\multicolumn{1}{c|}{e\_PM\_y} &
		\multicolumn{1}{c|}{Age} &
		\multicolumn{1}{c|}{E(B-V)} &
		\multicolumn{1}{c|}{$(m-M)_0$} &
		\multicolumn{1}{c|}{[Fe/H]}&
		\multicolumn{1}{c|}{e\_[Fe/H]}\\
		{[hms]}&{[dms]}&{[km/s]}&{[km/s]}&{[mas/yr]}&{[mas/yr]}&{[mas/yr]}&{[mas/yr]}&{[Gyr]}&{}&{}&{[dex]}&{[dex]}\\
		\hline
		08:51:23.4&+11:48:54&33.806&0.528&-5.906&0.951&-7.175&1.268&3.75-4.00&0.023&9.64&0.08&0.04\\
		\hline
		
	\end{tabular}
	\label{tab:cluster_par}
\end{table*}

\section{Results}
\label{sec:res}

Before comparing the [C/N] abundances of our sample of cluster members with theory, we need to assess the precision of the DR12/ASPCAP measurements. 
\citet{holtzman2015} discussed in detail checks for homogeneity of the derived abundances within individual clusters in their sample, and their agreement with previously measured values.
Regarding the first point, they examined the DR12/ASPCAP abundances under the assumption that clusters are chemically homogeneous for all elements except carbon and nitrogen, which are expected to have variations in giants because of surface mixing of CN-processed material. Any trend of chemical abundance versus $T_{eff}$ (that appears for example for Fe, Si, Ti) has been corrected for, except for C and N.   
Regarding the agreement with previously measured values, \citet{holtzman2015} conclude that there may be non-negligble systematic offsets betweeen DR12/ASPCAP [X/H] abundances and literature abundances for some elements at the level of 0.1-0.2~dex. As suggested by  \citet{holtzman2015}, we use in our work [X/Fe] abundances by subtracting [Fe/H] from the [X/H] abundances. This should at least partially cancel out possible trends and offsets in the results.

It is therefore crucial to test the CN abundances used in our analysis against independent, high-resolution spectroscopic measurements. 

\subsection{Comparison with high-resolution spectroscopy}
\label{sec:obs}

\begin{figure}
	\includegraphics[scale=0.39]{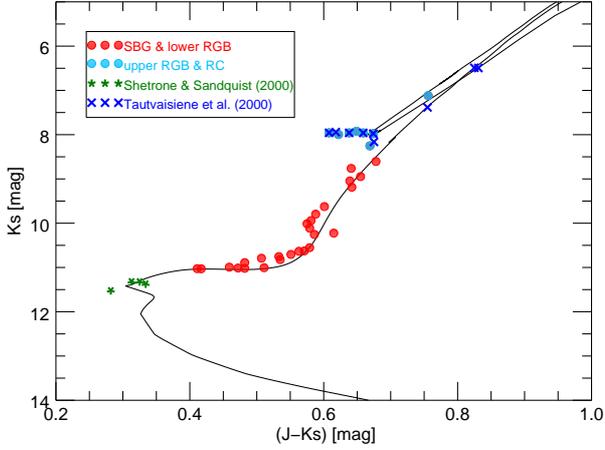}
	\caption{The plot shows our selection of M67 members from APOGEE together with stars analysed in other works. In particular, we consider the turn-off stars studied by \citet{shetrone2000} and the giant and clump stars from \citet{tautv2000}.}
	\label{fig:cn_iso_lit}
\end{figure}

\begin{figure}
	\includegraphics[scale=0.39]{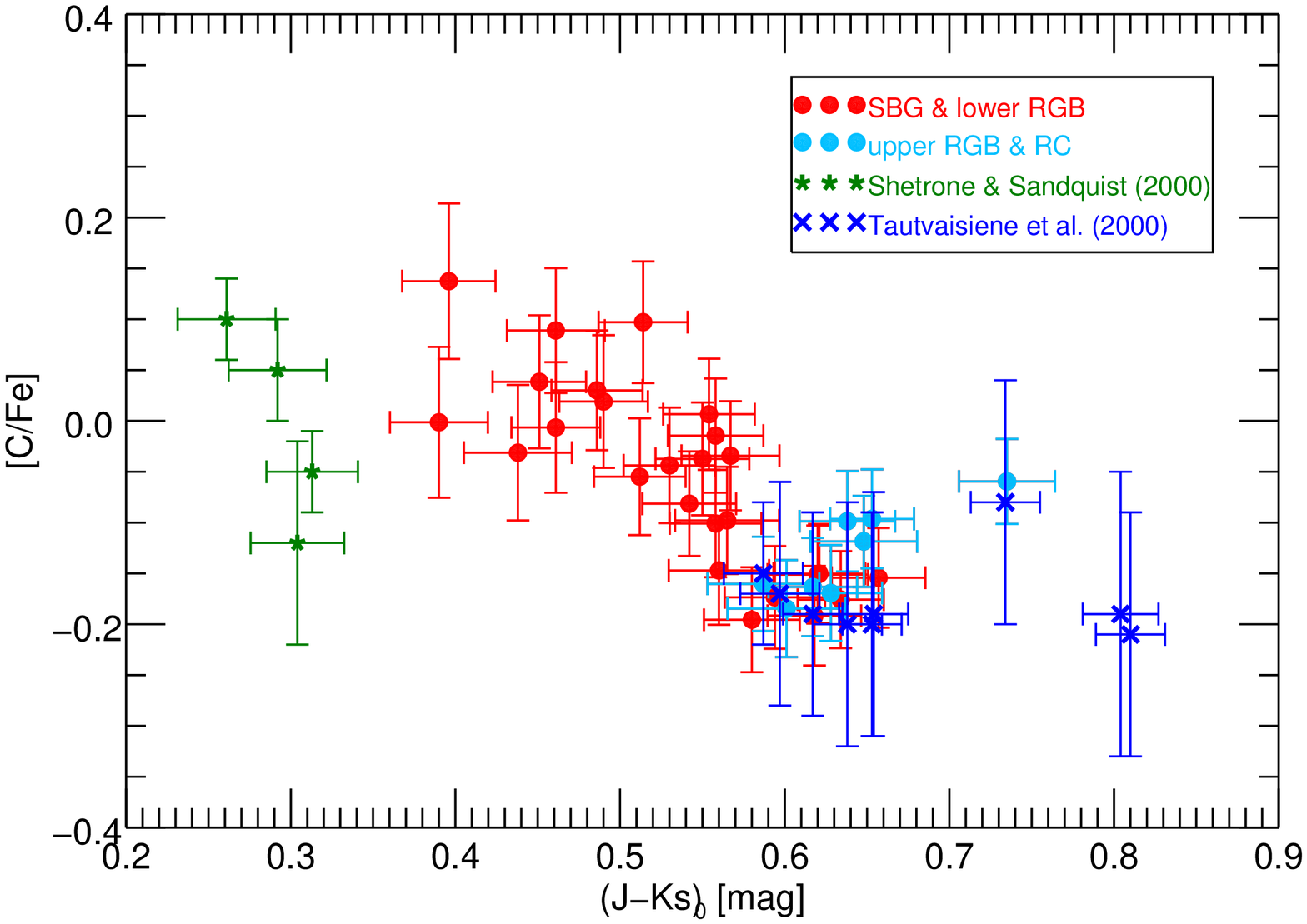}
	\includegraphics[scale=0.39]{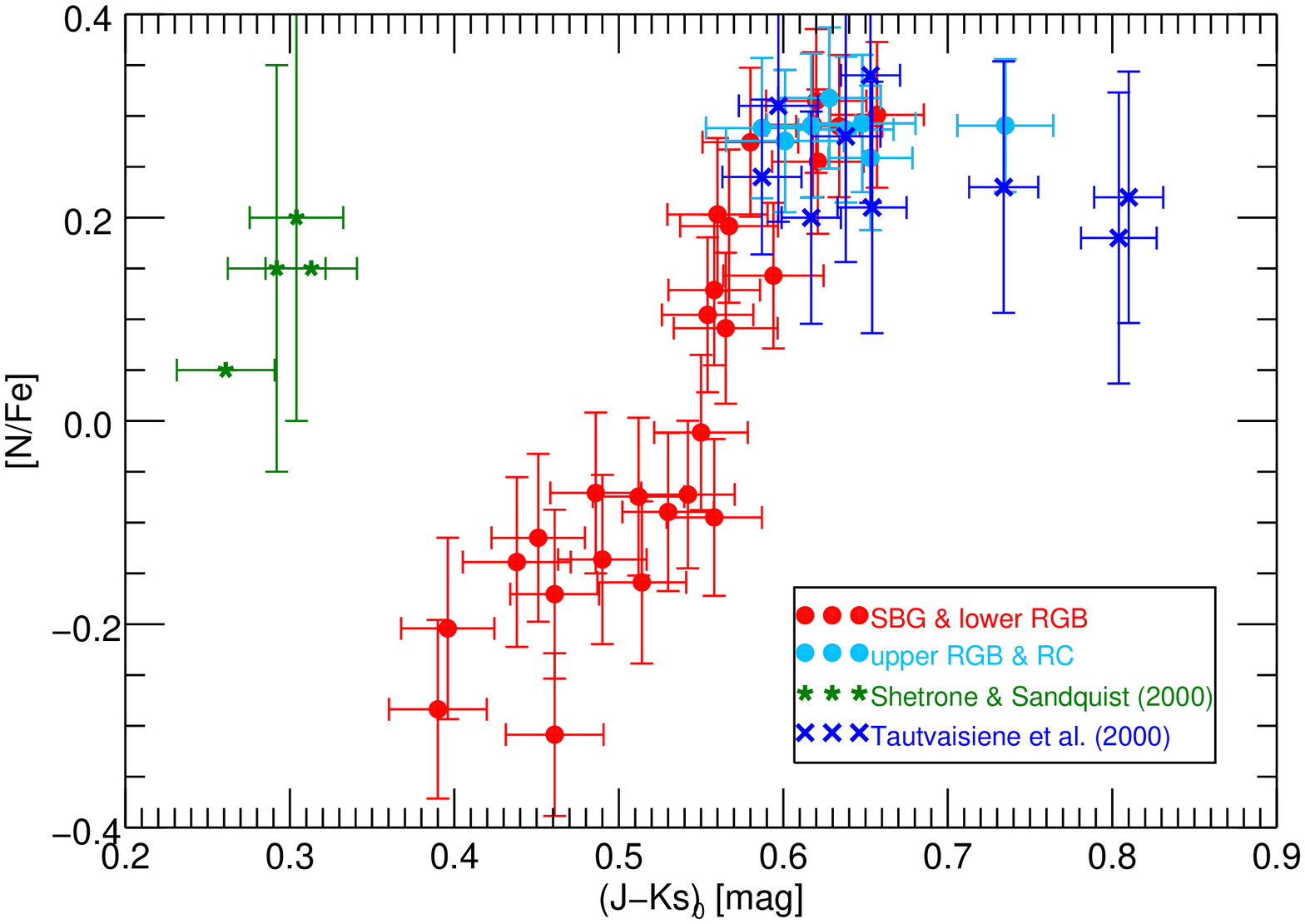}
	\includegraphics[scale=0.39]{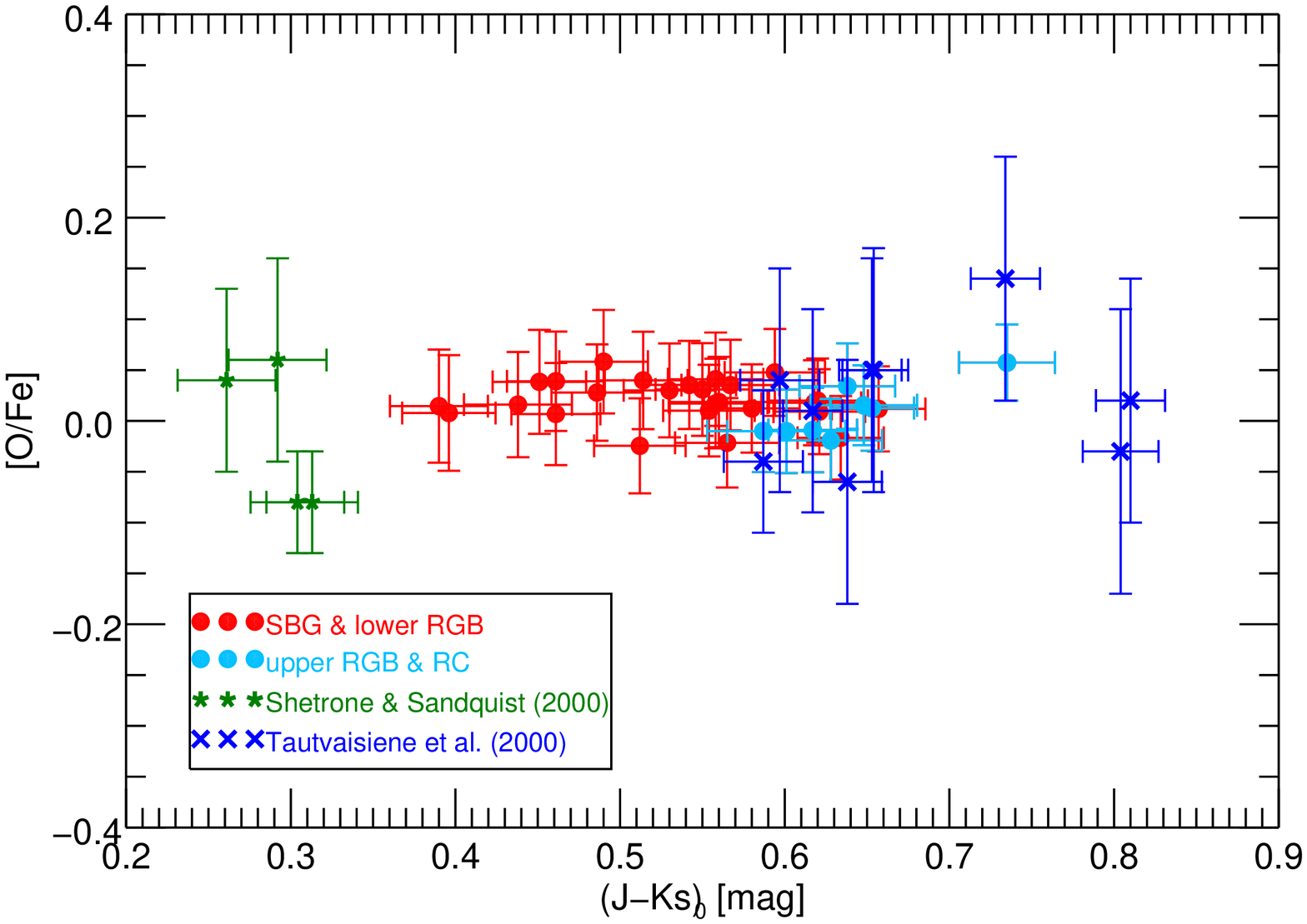}
	\caption{The [C/Fe], [N/Fe], and [O/Fe] abundances of the APOGEE sample and those from the literature are plotted as a function of colour. This picture confirms that the abundances from \citet{shetrone2000} and \citet{tautv2000} agree well with the expectation for solar values before the first dredge-up and with the values derived with DR12/ASPCAP after the first dredge-up. While [C/Fe] and [O/Fe] for the APOGEE sample agree very well with the literature values, [N/Fe] appears to be depleted by $\sim0.2$ dex on the subgiant branch.}
	\label{fig:cfe_comp}
\end{figure}

We found that there are not many published high-resolution spectroscopic estimates of C and N in M67 stars that can help us. Our comparisons will make use of the abundances of turn-off stars determined by \citet{shetrone2000}, and the abundances of red clump and post bump red giant branch stars by \citet{tautv2000} (see Fig.~\ref{fig:cn_iso_lit} and Table~\ref{tab:lit} for details about the selected stars). Assuming that the surface abundances are unchanged between the turn-off and the start of the FDU, the results of \citet{shetrone2000} should match the subgiant pre-FDU DR121/ASPCAP abundances.
Fig.~\ref{fig:cfe_comp} shows the DR12/ASPCAP [C/Fe], [N/Fe] and [O/Fe] values as a function of colour $(J-K_s)_0$, compared to the two independent sets of abundance determinations mentioned above. The colours of the stars sampled  by \citet{shetrone2000} and \citet{tautv2000} are found by matching the coordinates of the stars with the 2MASS catalogue.
We consider also the oxygen abundances for reasons that will become clearer in the course of this discussion.

The average red clump [C/Fe], [N/Fe], and [O/Fe] abundance ratios determined by \citet{tautv2000} are equal to $-0.18\pm0.02$,  $0.26\pm0.06$ and $0.01\pm0.05$ dex (the error bar is the $\sigma$ dispersion around the mean), respectively. 
Within the associated errors they are consistent with the average [C/Fe]=$-0.15\pm0.04$, [N/Fe]=$0.29\pm0.02$, and [O/Fe]=$0.00\pm0.02$ dex, which we obtain from the DR12/ASPCAP abundances of the RC. Notice that the colours (and $T_{eff}$) of red clump stars overlap with the colours of red giant branch stars at the FDU completion.

Regarding the turn-off abundances from \citet{shetrone2000}, the average [C/Fe] and [O/Fe] abundance ratios are equal to $-0.01\pm$0.10 and $-0.02\pm$0.08 dex. These values do  compare well, within the 1$\sigma$ errors, with the DR12/ASPCAP averages [C/Fe]=0.04$\pm$0.07 dex and [O/Fe]=0.02$\pm$0.02 dex obtained for pre-FDU objects (objects with $(J-K_s)_0<$0.47).
Assessing whether this consistency between turn-off and more evolved subgiant branch values  is a confirmation of the accuracy of DR12/ASPCAP abundances also for the hotter stars in the M67 sample requires a brief discussion about the effect of atomic diffusion on the surface abundances of these objects. 
Atomic diffusion (which usually denotes the combined effect of gravitational settling, thermal and chemical diffusion, and radiative levitation) can alter the surface abundances of M67 stars during the main sequence phase (causing either an increase or a decrease, 
depending on the ratio between local gravity and radiative acceleration of the various ions). The maximum variation 
compared to the initial abundances is reached around the turn-off. During the following subgiant branch phase 
the deepening of surface convection slowly restores the surface values back to essentially the initial abundances \citep[see, e.g.,][and references therein]{cs13}.

\citet{michaud2004} have calculated models and isochrones for M67 stars including the effect of atomic diffusion. Considering for example the surface Mg abundance, which behaves like O \citep{michaud2004}, these models predict a decrease during the main sequence phase, reaching a minimum around the turn-off, about 0.10~dex lower than the initial value, and then an increase along the subgiant branch until the surface abundances are restored to almost the initial values along the red giant branch. 
On the other hand also the Fe abundance has a similar behaviour, reaching a minimum at the turn-off about 0.08~dex lower that the initial value. Abundance ratios with respect to Fe would therefore display an even smaller variation along these evolutionary phases. 

\citet{onehag2014} presented a differential chemical abundance study of turn-off and main-sequence stars relative to hot subgiant 
branch stars in M67 (they include C, O, and Fe among other elements, but not N), which showed differences of their various measured [X/H] abundance ratios of the order of 0.02~dex, suggesting that diffusion is active (or maybe partially inhibited by some competing mechanism) although the evidence is not compelling, as they discussed. 
Their sample of subgiants (about 500~K hotter than the hotter DR12/ASPCAP subgiants) has average values of [C/Fe]=-0.04$\pm$0.05 dex and [O/Fe]=-0.04$\pm$0.07 dex, which within the errors agree with the DR12/ASPCAP mean subgiant pre-FDU abundances.
As a conclusion, there is no strong evidence for any sizeable variation of [C/Fe] and [O/Fe] between turn-off and the onset of the FDU. The average [C/Fe] and [O/Fe] ratios determined by \citet{shetrone2000} for a sample of turn-off stars, and those determined by \citet{onehag2014} for a sample of subgiant branch stars (hotter than DR12/ASPCAP ones) are both consistent within the errors with DR12/ASPCAP pre-FDU abundances. 
On the basis of these comparisons with red clump, turn-off and hot subgiant branch abundances from independent high-resolution spectroscopy, there is no indication that DR12/ASPCAP [C/Fe] and [O/Fe] abundances pre- and post-FDU need zero point and/or temperature dependent corrections.

However, the situation for [N/Fe] in the DR12/ASPCAP pre-FDU sample is different. It is clear from Fig.~\ref{fig:cfe_comp} that before the onset of the FDU [N/Fe] is not constant along the subgiant branch, as expected from the behaviour of [C/Fe] and [O/Fe], 
but rather displays a strong trend with colour (hence $T_{eff}$), i.e., [N/Fe] decreases steadily with decreasing colour along the SGB. The minimum [N/Fe] is equal to $\sim-$0.3~dex, a large depletion compared to the solar ratio.
The abundances of \citet{shetrone2000} are not very accurate and for two objects only upper limits could be determined. Nevertheless,  their data seem to be ruling out the low DR12/ASPCAP abundances (\citealt{onehag2014} do not measure N abundances).
In addition, it is hard to imagine a physical mechanism that changes only the surface N abundance along the subgiant branch, without 
modifying both C and O.
Finally, given the approximately solar metallicity of the cluster, and the scaled-solar values of C and O, it is very plausible to assume 
also a uniform solar [N/Fe] for the pre-FDU abundances. A value [N/Fe]=0.0 before the FDU would also be consistent with the constraint posed by the abundances of \citet{shetrone2000}.

In support of this interpretation of the DR12/ASPCAP results we refer to \citet{masseron2015}. Studying field subgiant stars in the thin and thick disc within APOGEE, they found an offset of $-0.2$ dex in the [N/Fe] abundance at solar metallicity, similar to what we observe in our selection of member stars for M67. 
This suggests that we are dealing with a systematic problem in the APOGEE data, or of the DR12/ASPCAP pipeline, and not with a peculiarity of M67. What still remains to be clarified is whether this problem is confined to the subgiant branch or whether it affects all stars within APOGEE. Due to the change of the surface C and N during and after the SGB, it is more difficult to determine deviations of the observed abundances from the predicted ones for field stars of which we do not know the mass. \citet{masseron2015} do not investigate the problem further and on the basis of the observed depletion in the SGB they shift the [N/Fe] abundances of the entire sample by $+0.2$ dex. 
While for the investigation of \citet{masseron2015} this does not affect the results, for our study it is important to
investigate whether this systematic offset is real.

We have shown that the [N/Fe] abundances of our sample of red giant branch and red clump stars fit very well those of stars independently observed and analysed by \citet{tautv2000}. Thus, we do not see any reason why the [N/Fe] abundances should be shifted by 0.2 dex for the entire sample of APOGEE stars. We claim that the reason for the systematic offset of [N/Fe] abundances in the APOGEE subgiant regime is due to the increasing temperature of the stars going from red to bluer colours. 
The [N/Fe] abundances in the DR12/ASPCAP pipeline are calculated from CN molecular lines. These become progressively weaker and more difficult to analyse as the effective temperature of the stars increases, thus leading to untrustworthy results in the subgiant region.  We therefore will not consider stars bluer than $(J-Ks)_0=0.54$ mag (the approximate value at which the APOGEE [N/Fe] abundances reach the solar value), corresponding to a temperature hotter than $T_{eff}\sim 5000$ K.

\subsection{Comparison with the models}
\label{sec:mod}

After having found a reasonable explanation for the offset of [N/Fe] abundances in the subgiant branch in DR12, we can proceed with the comparison between the APOGEE data and the predictions from stellar evolution models.

\begin{figure}
	\includegraphics[scale=0.39]{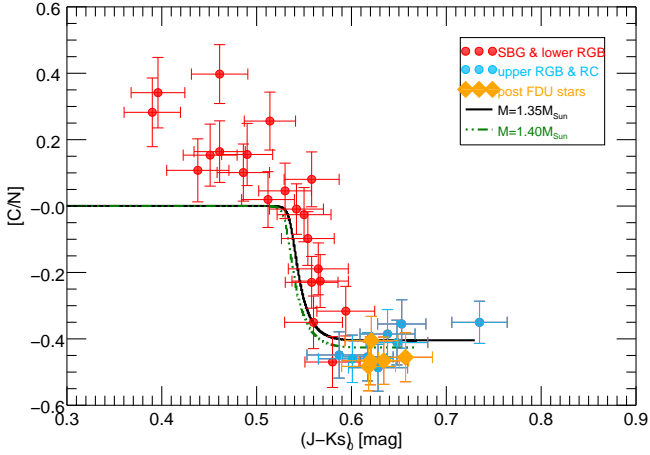}
	\caption{The plot shows the [C/N] abundances derived by DR12/ASPCAP as a function of the $(J-K_s)_0$ colour compared with two models  characterised by different masses (solid black line for $1.35 M_{\odot}$ and green dash-dotted line for $1.40 M_{\odot}$). The light blue circles represent stars of the RGB and RC that are excluded from the computation of the post-dredge-up [C/N] abundance due to possible extra mixing in their interior. The orange diamonds are the stars used to calculate the mean value for the post-dredge-up [C/N] abundance.}
	\label{fig:cn_mod}
\end{figure}

In Fig.~\ref{fig:cn_mod}, we plot the [C/N] abundance obtained for our sample of M67 members as a function of  $(J-K_s)_0$ colour together with two different models for stars of the masses $1.40 M_{\odot}$  and $1.35 M_{\odot}$, $\text{[Fe/H]}=0.06$ dex (from the BaSTI database). These correspond to an age of 4 and 3.75 Gyr, respectively, considered to be the turn-off age range of M67 following \citet{bellini2010}, who also employed the same BaSTI models used in this work.
The light-blue circles represent stars that might have undergone extra mixing and are therefore not suitable for the computation of the post-FDU [C/N] abundance. Finally, the orange diamonds represent the stars that we used to compute the post-FDU mean [C/N] abundance, as we will explain later in this section. We selected them based on the colour at which the FDU is complete, $(J-K_s)_0>0.6$ mag, and excluding stars brighter than the red giant bump luminosity from the models, $K_s<8.5$ mag.

\begin{figure}
	\includegraphics[scale=0.39]{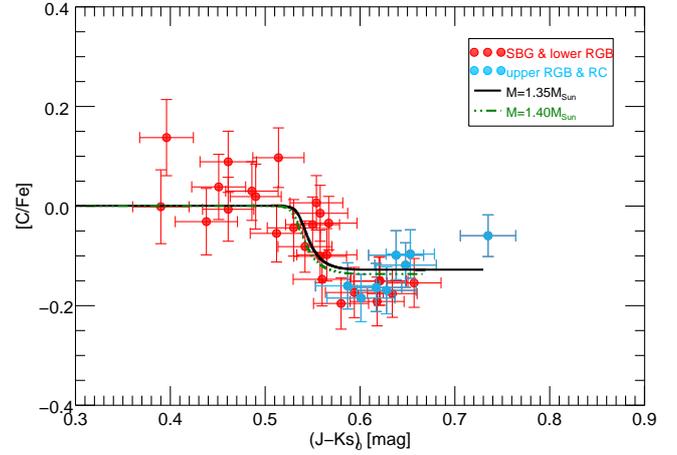}
	\includegraphics[scale=0.39]{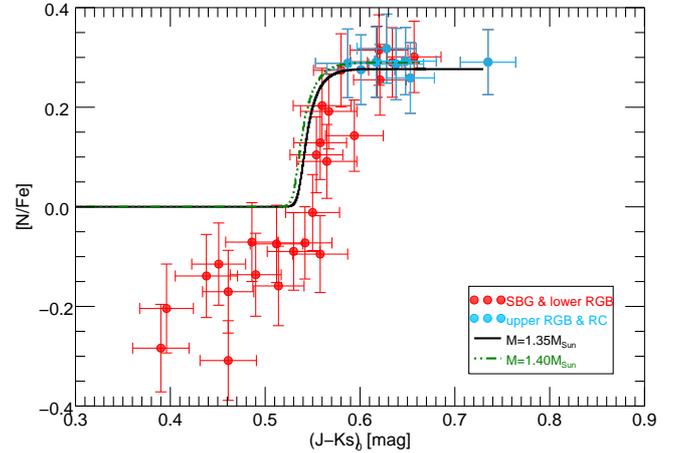}
	\caption{[C/Fe] and [N/Fe] as a function of the $(J-K_s)_0$ colour is plotted together with the respective model predictions. While the MS abundance of [C/Fe] derived with DR12/ASPCAP is consistent with a solar composition and with the models, the [N/Fe] abundance before the FDU is depleted and does not agree with the models.}
	\label{fig:cn_mod_sep}
\end{figure}

The DR12/ASPCAP abundances for stars with $(J-K_s)>0.54$ mag are consistent with the predictions of the models representing the age range of M67. In Fig.~\ref{fig:cn_mod_sep} the [C/Fe] and [N/Fe] abundances of the M67 stars and the respective models are plotted separately as a function of the $(J-K_s)_0$ colour. Also in this case the data are in good agreement with the models, if we exclude the [N/Fe] depletion in the subgiant range. We also compared the DR12/ASPCAP results with models using different assumption regarding envelope overshooting in the subgiant branch. We found that all models are consistent with the observational data and that we cannot distinguish between them within the errors (see \citealt{salaris2015} for details about the effect of envelope overshooting on $\text{[C/N]}_{FDU}$).

\begin{figure}
	\includegraphics[scale=0.43]{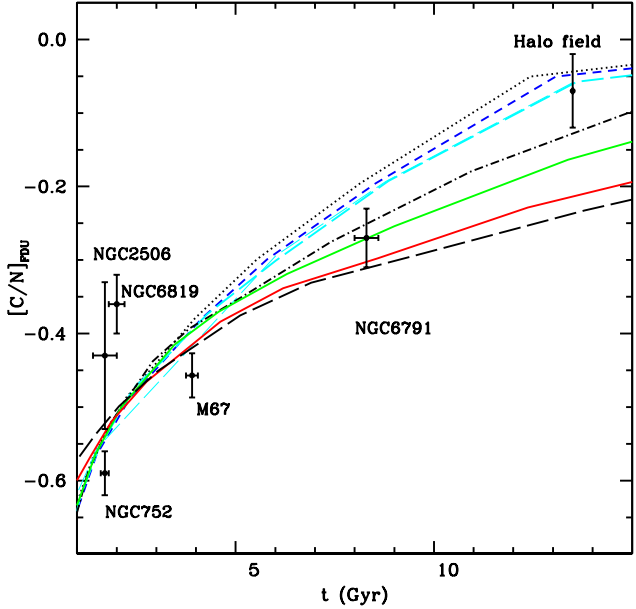}
	\caption{As Fig. 2 from \citet{salaris2015}: The plot shows the theoretical value of $\text{[C/N]}_{FDU}$ as a function of age for different metallicities. The curves shown computed for different [M/H]: -2.27 dex (black dotted line), -1.49 dex (blue short dashed), -1.27 dex (light blue long dashed), -0.66 dex (black dot-dashed), -0.35 dex (green solid), 0.06 dex (red solid) and 0.26 dex (black long dashed), respectively. Values for the $\text{[C/N]}_{FDU}$ of several Galactic open clusters and halo field stars available in the literature are also plotted (see \citealt{salaris2015} for details), together with the new $\text{[C/N]}_{FDU}$ value for M67 obtained in the current work.}
	\label{fig:sal_mod}
\end{figure}

We use five stars to determine the mean [C/N] abundance in M67 after the FDU (orange diamonds in Fig.~\ref{fig:cn_mod}), which we then compare with the models from \citet{salaris2015} for the age estimation of field stars (see Fig.~\ref{fig:sal_mod}). We obtain an abundance of $\text{[C/N]}_{FDU}=-0.46\pm 0.03$ dex, which is consistent with the prediction of $\text{[C/N]}_{FDU}=-0.4$ dex for the models with stellar mass $M=1.35M_{\odot}$ and of $\text{[C/N]}_{FDU}=-0.43$ dex for $M=1.4M_{\odot}$. 

Fig.~\ref{fig:sal_mod} updates Fig. 2 in \citet{salaris2015} with the $\text{[C/N]}_{FDU}$ of M67 obtained in this study. The errors are considerably smaller and the post-FDU $\text{[C/N]}_{FDU}$ is in better agreement with the models.

\subsection{Post-FDU extra mixing}
\label{sec:extra}

As mentioned above, we exclude bright RGB stars and RC stars to avoid problems related to possible extra mixing episodes that set in after the RGB bump. The mean molecular weight discontinuity left over by the FDU is expected to inhibit any extra-mixing below the convective envelope, but when this discontinuity is erased after the RGB bump, extra-mixing processes are possible, as shown for example by the evolution of the $\text{C}^{12}/\text{C}^{13}$ ratio in field giants at various metallicities \citep[][and references therein]{charbonnel2000}.These extra mixing processes can also decrease the [C/N] ratio compared to the FDU value (N increases
and C decreases), as seen, e.g., in field halo stars \citep{gratton2000}.

\begin{figure}
	\includegraphics[scale=0.39]{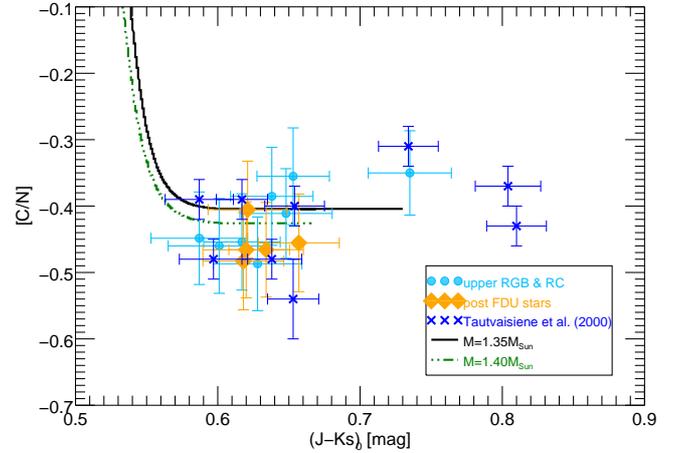}
	\caption{The plot shows the  [C/N] abundances for the post-FDU (orange diamonds),  and upper RGB and RC (light-blue circles) APOGEE stars in M67, as well as for the RGB and RC stars from \citet{tautv2000}. The data are compared to the models for stellar masses of $1.35 M_{\odot}$ (black solid line) and $1.4 M_{\odot}$ (green dashed-dotted line).}
	\label{fig:cn_mix}
\end{figure}

Our sample of M67 members contains seven red clump stars as well as one bright red giant branch (or possibly AGB) star. Fig.~\ref{fig:cn_mix} shows that the [C/N] abundances of the red clump stars are consistent with the post-FDU [C/N] abundance and do not appear to be affected by any extra mixing process. On the other hand, the [C/N] of the upper RGB  star ($K_s<8.5$ mag) is higher than $\text{[C/N]}_{FDU}$ and the value predicted by the models, contrarily to the drop in [C/N] expected from extra mixing. If we add the upper RGB and RC stars independently observed and analysed by \citet{tautv2000} to this sample, we see this picture confirmed (see also Fig 6 in \citealt{tautv2000}): RC stars have [C/N] abundances comparable to the $\text{[C/N]}_{FDU}$ value, while the [C/N] in upper RGB stars is slightly higher (see Fig.~\ref{fig:cn_mix}). This suggests that at the age and metallicity of M67 extra mixing after the red giant bump is not playing a significant role and that there might even be a different process taking place that has the opposite effect on the surface abundances of upper RGB stars.

We note that the study by \citet{souto2016}, analysing the chemical composition of six RC and six lower RGB stars in the $\sim2$ Gyr old open cluster NGC 2420 from APOGEE spectra found similar results. The RC stars in NGC 2420 show the same $^{13}\text{C}/^{14}\text{N}$ as the lower RGB stars, suggesting that no extra-mixing is taking place after the luminosity bump. 

Unfortunately, our sample of upper RGB stars (also taking into account the stars from \citealt{tautv2000}) is very small, so that no statistically significant conclusion can be drawn. We argue that more high-resolution spectroscopic observations of upper RGB stars in M67 are necessary to assess the role of extra mixing in such stars.

\section{Conclusions}

In this study we analyse the pre- and post-FDU [C/N] abundance in the old open cluster M67 and use it as a calibrator for the age-dating of field stars \citep[see][]{salaris2015}. In order to assess the accuracy of our analysis, we compare data obtained within the APOGEE DR12 survey and analysed by the stellar parameters and abundances pipeline ASPCAP with [C/Fe] and [N/Fe] values from the literature for turn-off \citep{shetrone2000}, red giant branch, and red clump \citep{tautv2000} stars in M67. 

We find the [C/Fe] abundances from DR12/ASPCAP to be in very good agreement with the literature, both before and after the FDU. Less straightforward are the results for [N/Fe]. While the values of our sample after the FDU are consistent within the errors with the results from \citet{tautv2000}, the abundances in the subgiant region appear systematically depleted and even show a trend with colour. Taking into account the results by \citet{masseron2015}, we assert that the depletion of [N/Fe] in the subgiant region is a problem common to the entire sample of APOGEE stars, and not a characteristic of M67. Possibly, the offset shown by the data is a result of the decreasing precision of the [N/Fe] measurements in DR12/ASPCAP with increasing temperature due to the weakening of the molecular bands used for the abundance determination. 

The problem of the systematic underestimation of [N/Fe] in the sub-giant branch seems to have been overcome in the recently published SDSS DR13 (see Appendix). This, however, happened on the expense of a much larger scatter in the overlall [N/Fe], [C/Fe], and [O/Fe] abundances. In addition, the post-FDU [N/Fe] abundance is not consistent any more with the other spectroscopic studies described in this work and with the evolutionary models (see Appendix for details).

Finally, we compare the DR12/ASPCAP [C/N] abundances with stellar evolutionary models which include core-overshooting and find that the observational data match the theoretical predictions very well within the errors when taking into account the findings described in the above paragraph. We obtain a new estimate for the mean post-FDU [C/N] in M67 of $\text{[C/N]}_{FDU}=-0.46\pm 0.03$ which is in good agreement and thus provides a stringent observational test of the $\text{[C/N]}_{FDU}-t-\text{[M/H]}$ relation of \citet{salaris2015} (see Fig.~\ref{fig:sal_mod} as an updated version of Fig. 2 in \citealt{salaris2015}).

\section*{Acknowledgements}

This work was supported by Sonderforschungsbereich SFB 881 "The Milky Way System" (subproject B5) of the German Research Foundation (DFG). The authors thank the anonymous referee, Jon Holtzman, Ricardo Schiavon, Sabine Reffert, and Corrado Boeche for helpful discussions. Maurizio Salaris acknowledges support from the SFB visitor programme. This work has made use of BaSTI web tools (http://albione.oa-teramo.inaf.it).

Funding for SDSS-III has been provided by the Alfred P. Sloan Foundation, the Participating Institutions, the National Science Foundation, and the U.S. Department of Energy Office of Science. The SDSS-III web site is http://www.sdss3.org/.

SDSS-III is managed by the Astrophysical Research Consortium for the Participating Institutions of the SDSS-III Collaboration including the University of Arizona, the Brazilian Participation Group, Brookhaven National Laboratory, Carnegie Mellon University, University of Florida, the French Participation Group, the German Participation Group, Harvard University, the Instituto de Astrofisica de Canarias, the Michigan State/Notre Dame/JINA Participation Group, Johns Hopkins University, Lawrence Berkeley National Laboratory, Max Planck Institute for Astrophysics, Max Planck Institute for Extraterrestrial Physics, New Mexico State University, New York University, Ohio State University, Pennsylvania State University, University of Portsmouth, Princeton University, the Spanish Participation Group, University of Tokyo, University of Utah, Vanderbilt University, University of Virginia, University of Washington, and Yale University.



\bibliographystyle{mnras}
\bibliography{references} 




\appendix

\renewcommand\thefigure{A.\arabic{figure}}
\setcounter{figure}{0}
\renewcommand\thetable{A.\arabic{table}}

\newpage
\subsection*{APPENDIX}

On 31st July 2016, the 13th data release of the Sloan Digital Sky Survey was made public. We performed the same analysis as for the DR12 data described in this work in order to determine the  degree of agreement between the new results derived with the updated ASPCAP pipeline with the old ones.

\begin{figure}
	\includegraphics[scale=0.39]{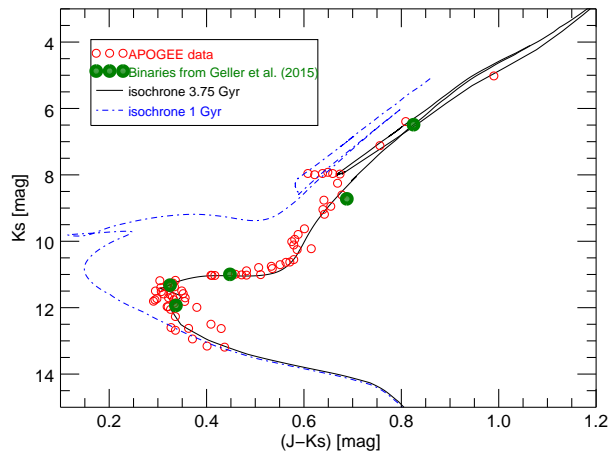}
	\caption{Selection of the members of M67 for the DR13. Binaries are highlighted in green and are excluded from further analysis. The 1 Gyr isochrone represents the age needed to fit the DR13 post-dredge-up [C/N] value with stellar evolutionary models.}
	\label{fig:iso_bin_dr13}
\end{figure}

In the DR13, some changes have been applied to the stellar parameter pipeline of APOGEE (see \citealt{sdss2016}, Holtzman et al. in prep.). For instance, different synthetic grids are now used for giants and dwarfs, allowing rotation for the latter. As a consequence, many turn-off and upper-main-sequence stars of M67, which had been previously excluded from our membership analysis, present now calibrated parameters from the ASPCAP pipeline. We thus find 78 member stars for M67, in contrast to the 34 found from the DR12 (see Fig.~\ref{fig:iso_bin_dr13}).

\begin{figure}
	\includegraphics[scale=0.39]{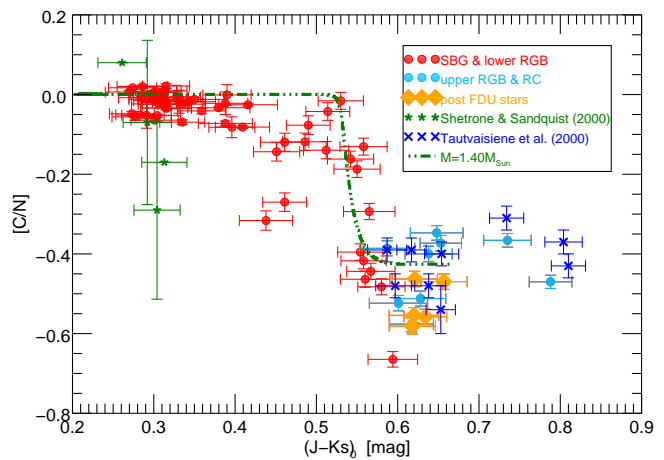}
	\caption{[C/N] abundance as a function of colour $(J-Ks)_0$. The orange diamonds indicate the stars that have completed the first dredge-up.}
	\label{fig:cn_dr13}
\end{figure}

Plotting the [C/N] abundance as a function of $(J-Ks)_0$ with the DR13 data, we find that the main-sequence and turn-off stars are consistent with a solar chemical composition. The sub-giant and giant stars, on the other hand, show a very large scatter and the post-dredge-up average [C/N] value for the cluster is lower with respect to the DR12 and the high-resolution spectroscopic data found in \citet{shetrone2000} and \citet{tautv2000} (see Fig.~\ref{fig:cn_dr13}).

\begin{figure}
	\includegraphics[scale=0.39]{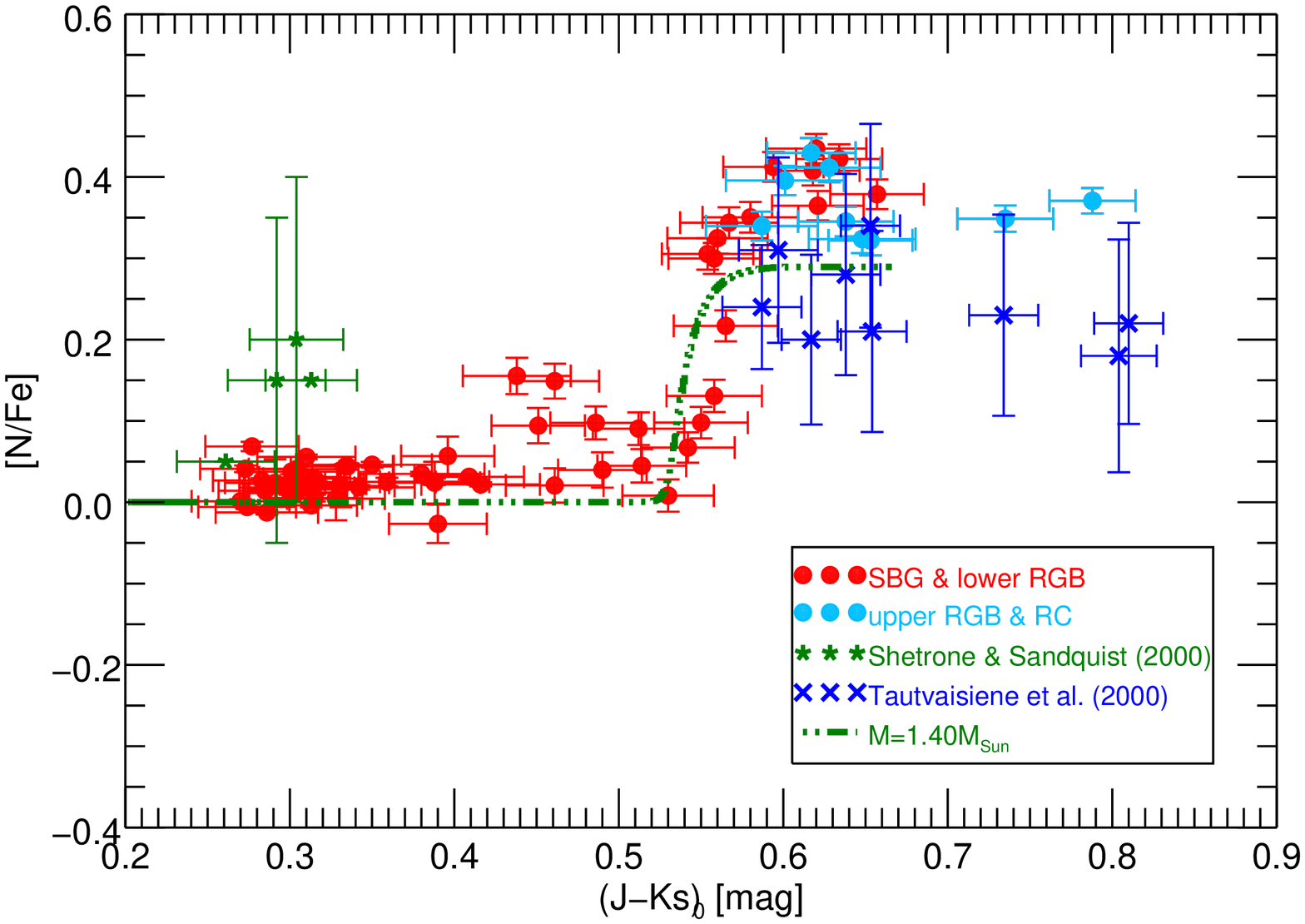}
	\includegraphics[scale=0.39]{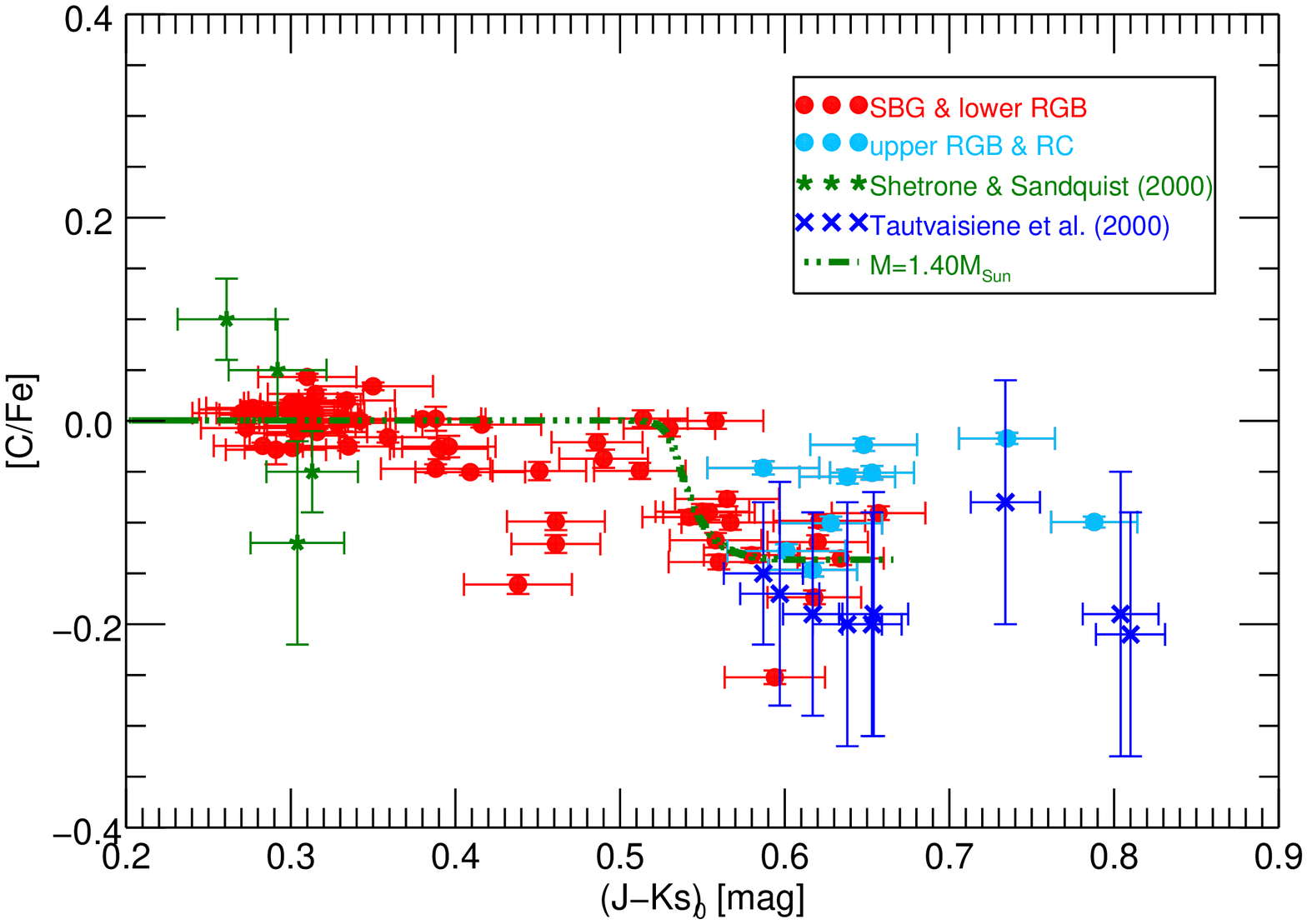}
	\includegraphics[scale=0.39]{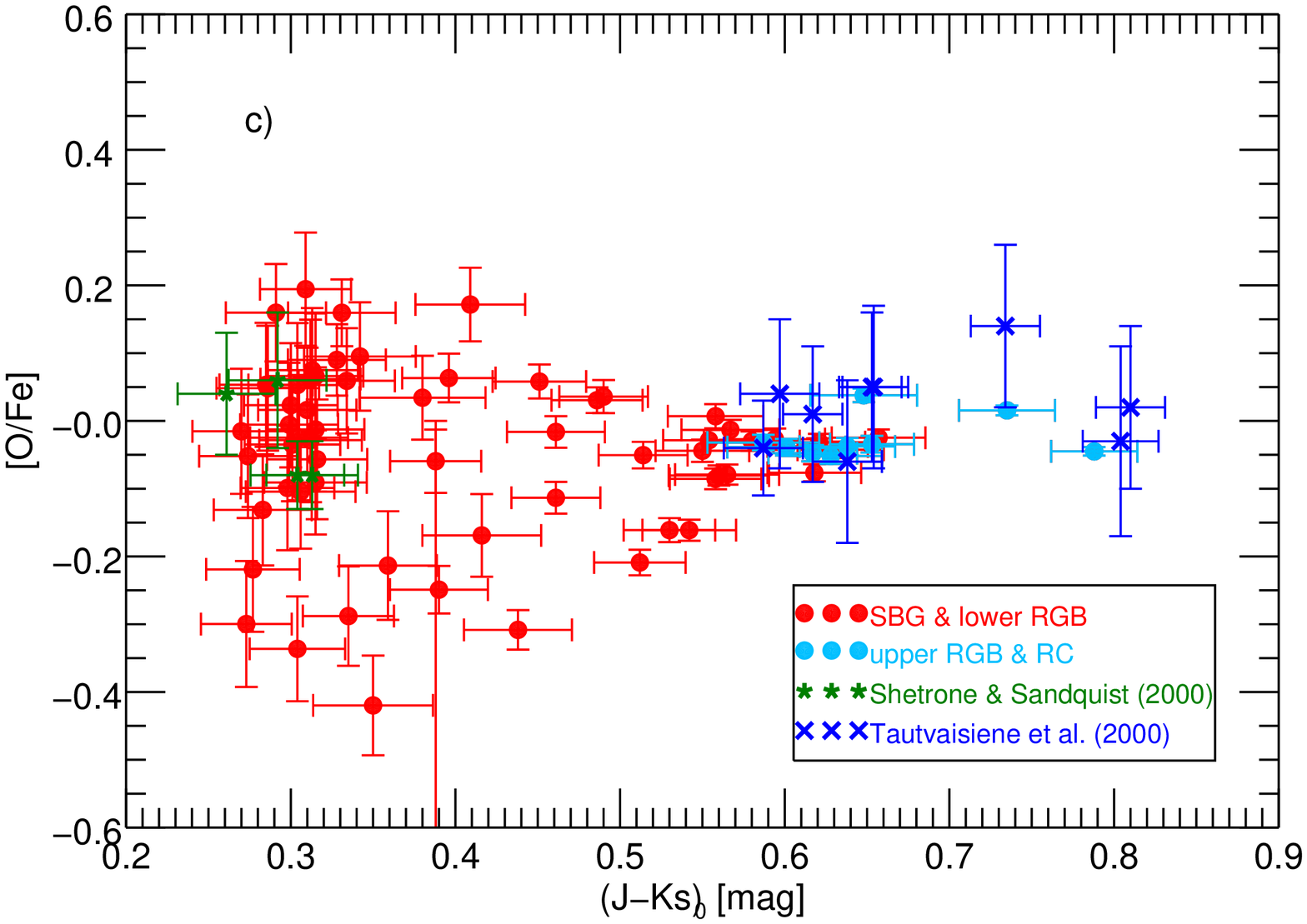}
	\caption{[N/Fe], [C/Fe], and [O/Fe] abundance as a function of colour $(J-Ks)_0$.}
	\label{fig:nfe_dr13}
\end{figure}

When looking at the different elemental abundances separately, the changes applied to the ASPCAP pipeline in the DR13 seem to have solved the problem of the depletion for the [N/Fe] abundances in the sub-giant branch (see Fig.~\ref{fig:nfe_dr13}, panel a). Nevertheless, while giant-branch and red-clump stars in the DR12 were consistent with the results of high-resolution spectroscopic studies found in the literature, in the DR13 the post-dredge-up values for [N/Fe] are systematically enhanced with respect to \citet{tautv2000}. In particular, the red clump and upper red giant branch stars of APOGEE DR13 have on average a [N/Fe] value that is $0.12$ dex higher than in \citet{tautv2000}. 

Plotting the DR13 results for the [C/Fe] and [O/Fe] abundances as a function of $(J-Ks)_0$, we observe a very large scatter of the data affecting the giant stars for the former and the main-sequence and turn-off stars for the latter (see Fig.~\ref{fig:nfe_dr13}, panel b and c). In comparison with \citet{tautv2000} the [C/Fe] abundance in red clump and upper red giant stars in APOGEE DR13 is on average $0.10$ dex higher, while the [O/Fe] abundance is only $0.05$ dex lower.

In addition, we note that the stellar evolutionary models described in Sec.~\ref{sec:mod}, representing a stellar mass of $1.4 M_{\odot}$ and an age of 3.75 Gyr, which were consistent with the results of the DR12, reproduce very poorly the abundances resulting from the DR13 (see Fig.~\ref{fig:cn_dr13} and Fig.~\ref{fig:nfe_dr13}).
We run a $\chi^2$ test to quantify and compare the goodness of the fit between the APOGEE DR12 and 13 data, with
\begin{equation}
\chi^2=\sum_{i} \dfrac{(O_i-E_i)^2}{\sigma_i^2},
\end{equation}
where $O_i$ are the observed values of [C/N], $E_i$ the $\text{[C/N]}_{FDU}$ abundance predicted by the models, and $\sigma_i$ the errors of the observations. We considered only stars that have already completed the first dredge-up, i.e. with $(J-Ks)_0>0.6$ and $Ks>8.5$ mag (orange diamonds in Fig.~\ref{fig:cn_dr13}), and found $\chi^2=1.5$ for the DR12 versus $\chi^2=164.6$ for the DR13.\\
If, in addition, we wanted to fit the mean $\text{[C/N]}_{FDU}$ obtained from the same selection of stars for the DR13 ([C/N] $=-0.53$ dex) with the same set of stellar evolutionary models, we would obtain a stellar mass corresponding to an age of $\sim 1$ Gyr, which, however, is not consistent with the observed CMD of M67, as shown in Fig.~\ref{fig:iso_bin_dr13}.

To summarise, the changes applied to the ASPCAP pipeline in DR13 seem to have improved the problem of the systematic depletion in [N/Fe] observed in hot stars in the DR12. Nevertheless, the overall results for the abundances of sub-giant and giant stars in the DR13 are in worse agreement with literature values than in the DR12. This would indicate that further adjustments have to be made to the ASPCAP pipeline in order to improve the abundance determination for certain elements at cooler temperatures and higher luminosities.

\newpage
	\begin{landscape}
\begin{table}
	\caption{List of the selected M67 members with their coordinates, radial velocities from ASPCAP, proper motions from PPMXL, infrared colours from 2MASS, and the respective errors.}
\begin{tabular}{|r|r|r|r|r|r|r|r|r|r|r|r|r|r|r|}
	\hline
	\multicolumn{1}{|r|}{Nr} &
	\multicolumn{1}{r|}{RA (J2000)} &
	\multicolumn{1}{r|}{Dec (J2000)} &
	\multicolumn{1}{r|}{RV} &
	\multicolumn{1}{r|}{RV\_err} &
	\multicolumn{1}{r|}{PMx} &
	\multicolumn{1}{r|}{PMy} &
	\multicolumn{1}{r|}{PM\_tot} &
	\multicolumn{1}{r|}{PM\_err} &
	\multicolumn{1}{r|}{Jmag} &
	\multicolumn{1}{r|}{Hmag} &
	\multicolumn{1}{r|}{Ksmag} &
	\multicolumn{1}{r|}{e\_Jmag} &
	\multicolumn{1}{r|}{e\_Hmag} &
	\multicolumn{1}{r|}{e\_Ksmag} \\
	{}&{[hms]}&{[dms]}&{[km/s]}&{[km/s]}&{[mas/yr]}&{[mas/yr]}&{[mas/yr]}&{[mas/yr]}&{[mag]}&{[mag]}&{[mag]}&{[mag]}&{[mag]}&{[mag]}\\
	\hline
  1 & 08:50:36.134 & +11:43:18.005 & 34.087 & .010 & -6.270 & -5.136 & 8.105 & 1.000 & 11.131 & 10.644 & 10.552 & .020 & .022 & .021\\
  2 & 08:50:49.949 & +11:49:12.727 & 33.716 & .015 & -6.798 & -6.952 & 9.724 & 1.100 & 11.372 & 10.960 & 10.890 & .021 & .020 & .017\\
  3 & 08:50:58.162 & +11:52:22.361 & 34.091 & .010 & -6.376 & -5.451 & 8.389 & 1.300 & 11.197 & 10.707 & 10.626 & .022 & .020 & .018\\
  4 & 08:51:08.390 & +11:47:12.142 & 33.509 & .008 & -6.597 & -6.657 & 9.372 & 1.000 & 10.691 & 10.195 & 10.112 & .022 & .022 & .017\\
  5 & 08:51:12.697 & +11:52:42.370 & 34.389 & .002 & -6.296 & -6.079 & 8.752 & 1.900 & 8.650 & 8.122 & 7.976 & .018 & .018 & .018\\
  6 & 08:51:15.641 & +11:50:56.162 & 34.423 & .021 & -4.453 & -6.784 & 8.115 & 1.000 & 11.485 & 11.094 & 11.013 & .022 & .019 & .018\\
  7 & 08:51:17.040 & +11:50:46.424 & 33.717 & .003 & -6.574 & -7.855 & 10.243 & 2.200 & 9.284 & 8.712 & 8.606 & .022 & .021 & .018\\
  8 & 08:51:18.773 & +11:51:18.673 & 34.360 & .012 & -5.032 & -7.796 & 9.279 & 1.100 & 11.502 & 11.089 & 11.020 & .022 & .020 & .020\\
  9 & 08:51:18.972 & +11:58:11.021 & 33.966 & .007 & -5.875 & -8.473 & 10.311 & 1.000 & 10.587 & 10.095 & 10.012 & .022 & .020 & .017\\
  10 & 08:51:21.564 & +11:46:06.121 & 34.850 & .003 & -4.703 & -6.857 & 8.315 & 2.000 & 9.602 & 9.085 & 8.947 & .019 & .020 & .018\\
  11 & 08:51:26.186 & +11:53:52.048 & 34.164 & .002 & -5.243 & -5.835 & 7.845 & 1.856 & 8.619 & 8.113 & 7.960 & .020 & .020 & .021\\
  12 & 08:51:28.988 & +11:50:33.072 & 33.450 & .002 & -7.116 & -6.483 & 9.626 & .900 & 8.566 & 8.072 & 7.958 & .024 & .018 & .024\\
  13 & 08:51:29.351 & +11:45:27.558 & 33.095 & .011 & -4.876 & -7.613 & 9.040 & 1.100 & 11.287 & 10.864 & 10.754 & .022 & .020 & .017\\
  14 & 08:51:35.403 & +11:57:56.455 & 33.449 & .017 & -5.298 & -8.643 & 10.138 & 1.000 & 11.447 & 11.143 & 11.030 & .021 & .022 & .019\\
  15 & 08:51:35.778 & +11:53:34.714 & 34.006 & .007 & -6.155 & -7.711 & 9.866 & 1.000 & 10.522 & 10.023 & 9.941 & .020 & .020 & .023\\
  16 & 08:51:38.626 & +12:20:14.183 & 33.755 & .012 & -6.782 & -8.363 & 10.767 & 1.200 & 11.298 & 10.866 & 10.791 & .021 & .019 & .018\\
  17 & 08:51:39.382 & +11:51:45.659 & 34.523 & .006 & -5.272 & -5.846 & 7.873 & .900 & 10.383 & 9.889 & 9.795 & .020 & .022 & .022\\
  18 & 08:51:42.342 & +11:50:07.631 & 34.349 & .004 & -4.789 & -8.241 & 9.531 & .900 & 9.829 & 9.339 & 9.187 & .022 & .023 & .017\\
  19 & 08:51:42.358 & +11:51:23.098 & 33.600 & .003 & -5.744 & -6.789 & 8.893 & .900 & 9.403 & 8.854 & 8.762 & .020 & .023 & .023\\
  20 & 08:51:43.888 & +11:56:42.580 & 33.018 & .002 & -5.453 & -7.740 & 9.468 & 1.867 & 8.618 & 8.114 & 7.996 & .018 & .034 & .031\\
  21 & 08:51:44.013 & +11:46:24.521 & 32.980 & .017 & -4.664 & -7.249 & 8.619 & 1.000 & 11.438 & 11.110 & 11.027 & .022 & .022 & .020\\
  22 & 08:51:44.741 & +11:46:46.031 & 33.593 & .011 & -4.112 & -8.268 & 9.234 & 1.100 & 11.357 & 10.918 & 10.822 & .021 & .022 & .017\\
  23 & 08:51:45.078 & +11:47:45.938 & 33.017 & .004 & -6.698 & -8.237 & 10.617 & 1.900 & 9.684 & 9.183 & 9.045 & .022 & .023 & .018\\
  24 & 08:51:48.838 & +11:56:51.191 & 34.398 & .011 & -6.823 & -8.132 & 10.615 & 1.100 & 11.256 & 10.779 & 10.705 & .022 & .022 & .017\\
  25 & 08:51:55.672 & +12:17:57.332 & 33.381 & .015 & -7.462 & -7.867 & 10.844 & 1.200 & 11.516 & 11.115 & 11.005 & .020 & .024 & .018\\
  26 & 08:51:56.117 & +11:50:14.777 & 34.614 & .010 & -6.320 & -7.828 & 10.060 & 1.200 & 11.197 & 10.726 & 10.634 & .022 & .020 & .018\\
  27 & 08:51:59.522 & +11:55:04.922 & 34.426 & .002 & -5.641 & -6.780 & 8.820 & .900 & 8.597 & 8.084 & 7.959 & .020 & .024 & .018\\
  28 & 08:52:10.973 & +11:31:49.152 & 33.892 & .002 & -6.773 & -5.520 & 8.737 & 1.600 & 8.921 & 8.388 & 8.252 & .027 & .071 & .018\\
  29 & 08:52:11.343 & +11:45:38.023 & 33.108 & .016 & -5.249 & -8.044 & 9.605 & 1.000 & 11.452 & 11.082 & 10.993 & .026 & .034 & .020\\
  30 & 08:52:16.564 & +11:19:38.017 & 33.769 & .001 & -7.710 & -5.206 & 9.303 & 1.532 & 7.875 & 7.233 & 7.119 & .023 & .036 & .018\\
  31 & 08:52:18.569 & +11:44:26.304 & 33.725 & .002 & -4.994 & -9.190 & 10.459 & .900 & 8.572 & 8.087 & 7.923 & .021 & .057 & .023\\
  32 & 08:52:20.030 & +11:27:36.252 & 33.946 & .007 & -4.690 & -3.923 & 6.114 & 1.900 & 10.839 & 10.383 & 10.253 & .026 & .032 & .018\\
  33 & 08:52:56.250 & +11:48:53.939 & 32.950 & .007 & -6.681 & -9.800 & 11.861 & 1.000 & 10.839 & 10.315 & 10.224 & .023 & .026 & .020\\
  34 & 08:53:46.727 & +11:23:30.714 & 33.098 & .005 & -7.282 & -6.598 & 9.827 & .900 & 10.225 & 9.730 & 9.624 & .019 & .026 & .022\\
	\hline\end{tabular}
\label{tab:members_par}
\end{table}
\end{landscape}

	\begin{landscape}
		\begin{table}
			\centering
			\caption{List of the ASPCAP abundances and their errors for our selection of M67 members.}
\begin{tabular}{|r|r|r|r|r|r|r|r|r|}
	\hline
	\multicolumn{1}{|r|}{Nr} &
	\multicolumn{1}{r|}{[Fe/H]} &
	\multicolumn{1}{r|}{[Fe/H]\_err} &
	\multicolumn{1}{r|}{[C/Fe]} &
	\multicolumn{1}{r|}{[C/Fe]\_err} &
	\multicolumn{1}{r|}{[N/Fe]} &
	\multicolumn{1}{r|}{[N/Fe]\_err} &
	\multicolumn{1}{r|}{[O/Fe]} &
	\multicolumn{1}{r|}{[O/Fe]\_err} \\
	{}&{[dex]}&{[dex]}&{[dex]}&{[dex]}&{[dex]}&{[dex]}&{[dex]}&{[dex]}\\
\hline
 1 & .080 & .034 & -.014 & .056 & -.095 & .077 & .041 & .046\\
 2 & .016 & .035 & -.006 & .064 & -.170 & .083 & .007 & .050\\
 3 & .059 & .033 & -.037 & .056 & -.011 & .076 & .031 & .046\\
 4 &  .121 & .033 & -.101 & .053 & .129 & .074 & .017 & .044\\
 5 & .079 & .032 & -.096 & .049 & .259 & .071 & .012 & .042\\
 6 & .066 & .035 & .038 & .065 & -.115 & .083 & .038 & .051\\
 7 & .086 & .033 & -.154 & .049 & .301 & .072 & .012 & .042\\
 8 & .071 & .034 & .089 & .061 & -.309 & .080 & .039 & .049\\
 9 & .029 & .034 & .007 & .055 & .104 & .076 & .010 & .045\\
 10 & .140 & .032 & -.176 & .048 & .290 & .070 & -.017 & .041\\
 11 & .059 & .033 & -.099 & .049 & .287 & .072 & .034 & .042\\
 12 & .118 & .032 & -.160 & .046 & .288 & .069 & -.010 & .040\\
 13 & .096 & .034 & -.055 & .057 & -.074 & .078 & -.024 & .047\\
 14 & .009 & .036 & .137 & .076 & -.204 & .089 & .008 & .057\\
 15 & .072 & .033 & -.147 & .053 & .203 & .075 & .019 & .044\\
 16 & .030 & .034 & .030 & .059 & -.071 & .079 & .028 & .047\\
 17 & .045 & .033 & -.034 & .054 & .192 & .075 & .035 & .045\\
 18 & .120 & .032 & -.150 & .049 & .255 & .071 & .009 & .042\\
 19 & .088 & .032 & -.151 & .048 & .315 & .071 & .020 & .041\\
 20 & .128 & .032 & -.185 & .048 & .275 & .070 & -.010 & .041\\
 21 & .043 & .035 & -.001 & .074 & -.284 & .088 & .015 & .056\\
 22 & .023 & .034 & .097 & .060 & -.159 & .080 & .040 & .048\\
 23 & .105 & .033 & -.192 & .049 & .291 & .071 & .018 & .042\\
 24 & .056 & .034 & -.044 & .057 & -.090 & .078 & .030 & .046\\
 25 & .009 & .035 & .019 & .065 & -.136 & .083 & .058 & .051\\
 26 & .157 & .033 & -.081 & .051 & -.072 & .073 & .035 & .043\\
 27 & .102 & .032 & -.163 & .048 & .291 & .071 & -.009 & .042\\
 28 & .113 & .032 & -.118 & .045 & .293 & .068 & .015 & .039\\
 29 & .068 & .035 & -.031 & .067 & -.139 & .083 & .016 & .052\\
 30 & .049 & .031 & -.060 & .042 & .291 & .065 & .057 & .038\\
 31 & .143 & .032 & -.169 & .047 & .318 & .069 & -.019 & .041\\
 32 & .110 & .033 & -.098 & .053 & .091 & .074 & -.022 & .044\\
 33 & .171 & .032 & -.174 & .050 & .143 & .072 & .048 & .043\\
 34 & .112 & .033 & -.196 & .052 & .274 & .073 & .012 & .043\\
	\hline\end{tabular}

\label{tab:members_ab}
		\end{table}
	\end{landscape}
	
\newpage
\begin{landscape}
			\begin{table}
					\caption{Stars from the literature used for comparison with our APOGEE/ASPCAP data. The stars from \citet{shetrone2000} follow the nomenclature from \citet{sanders1977}, those from \citet{tautv2000} follow \citet{fagerholm1906}. The proper motions are from \citet{frolov1986}, and the magnitudes from 2MASS. The errors of the carbon and oxygen abundances in \citet{tautv2000} are indeed lower limits to the errors, as the uncertainties for the [C/H] and [O/H] abundances were not provided and only the errors to the [Fe/H] abundances could be used in the computation.}
\begin{tabular}{|l|r|r|r|r|r|r|r|r|r|r|r|r|r|r|r|r|r|r|}
	\hline
	\multicolumn{1}{|l|}{Name} &
	\multicolumn{1}{r|}{RA (J2000)} &
	\multicolumn{1}{r|}{Dec (J2000)} &
	\multicolumn{1}{r|}{PMx} &
	\multicolumn{1}{r|}{PMy} &
	\multicolumn{1}{r|}{Jmag} &
	\multicolumn{1}{r|}{Hmag} &
	\multicolumn{1}{r|}{Kmag} &
	\multicolumn{1}{r|}{e\_Jmag} &
	\multicolumn{1}{r|}{e\_Hmag} &
	\multicolumn{1}{r|}{e\_Kmag} &
	\multicolumn{1}{r|}{[Fe/H]} &
	\multicolumn{1}{r|}{e\_[Fe/H]} &
	\multicolumn{1}{r|}{[C/Fe]} &
	\multicolumn{1}{r|}{e\_[C/Fe]} &
	\multicolumn{1}{r|}{[N/Fe]} &
	\multicolumn{1}{r|}{e\_[N/Fe]} &
	\multicolumn{1}{r|}{[O/Fe]} &
	\multicolumn{1}{r|}{e\_[O/Fe]} \\
		{}&{[hms]}&{[dms]}&{[mas/yr]}&{[mas/yr]}&{[mag]}&{[mag]}&{[mag]}&{[mag]}&{[mag]}&{[mag]}&{[dex]}&{[dex]}&{[dex]}&{[dex]}&{[dex]}&{[dex]}&{[dex]}&{[dex]}\\
	\hline
	\multicolumn{19}{c}{Stars from \citet{shetrone2000}}\\
\hline
S815 & 08:50:54.37 & +11:56:28.8 & -.003 & -.001 & 11.706 & 11.435 & 11.372 & .022 & .019 & .017 & -.050 & .040 & -.050 & .040 & .150 &  & -.080 & .050\\
S1183 & 08:51:49.93 & +11:33:18.5 & -.007 & -.000 & 11.632 & 11.372 & 11.319 & .022 & .022 & .020 & -.040 & .080 & .050 & .050 & .150 & .200 & .060 & .100\\
S1271 & 08:51:34.28 & +11:49:43.9 & -.004 & -.003 & 11.804 & 11.569 & 11.522 & .022 & .022 & .020 & -.070 & .030 & .100 & .040 & .050 &  & .040 & .090\\
S821 & 08:50:51.80 & +11:56:55.5 & -.001 & .001 & 11.646 & 11.382 & 11.321 & .022 & .020 & .018 & -.040 & .040 & -.120 & .100 & .200 & .200 & -.080 & .050\\
\hline
\multicolumn{19}{c}{Stars from \citet{tautv2000}}\\
\hline
F84 & 08:51:12.70 & +11:52:42.5 & -.006 & -.001 & 8.650 & 8.122 & 7.976 & .018 & .018 & .018 & -.020 & .110 & -.200 & .110 & .340 & .125 & .050 & .110\\
F105 & 08:51:17.11 & +11:48:15.9 & -.003 & -.004 & 8.140 & 7.526 & 7.385 & .027 & .018 & .021 & -.050 & .120 & -.080 & .120 & .230 & .124 & .140 & .120\\
F108 & 08:51:17.48 & +11:45:22.5 & -.003 & -.002 & 7.325 & 6.683 & 6.494 & .021 & .020 & .021 & -.020 & .120 & -.210 & .120 & .220 & .124 & .020 & .120\\
F141 & 08:51:22.81 & +11:48:01.7 & -.005 & -.006 & 8.560 & 8.075 & 7.942 & .023 & .033 & .024 & -.010 & .110 & -.170 & .110 & .310 & .114 & .040 & .110\\
F151 & 08:51:26.19 & +11:53:52.0 & -.006 & -.004 & 8.619 & 8.113 & 7.960 & .020 & .020 & .021 & .010 & .120 & -.200 & .120 & .280 & .124 & -.060 & .120\\
F164 & 08:51:28.99 & +11:50:32.9 & -.005 & -.003 & 8.566 & 8.072 & 7.958 & .024 & .018 & .024 & .000 & .070 & -.150 & .070 & .240 & .076 & -.040 & .070\\
F170 & 08:51:29.91 & +11:47:16.7 & -.003 & -.009 & 7.314 & 6.681 & 6.489 & .020 & .027 & .023 & -.020 & .140 & -.190 & .140 & .180 & .143 & -.030 & .140\\
F224 & 08:51:43.56 & +11:44:26.3 & -.003 & -.004 & 8.838 & 8.294 & 8.163 & .032 & .020 & .021 & -.110 & .120 & -.190 & .120 & .210 & .124 & .050 & .120\\
F266 & 08:51:59.54 & +11:55:05.2 & -.005 & -.007 & 8.597 & 8.084 & 7.959 & .020 & .024 & .018 & -.020 & .100 & -.190 & .100 & .200 & .104 & .010 & .100\\
			\hline\end{tabular}
\label{tab:lit}
			\end{table}
\end{landscape}


\bsp	
\label{lastpage}
\end{document}